\documentclass[aps, prl, showpacs, twocolumn, 10pt]{revtex4-1}
\usepackage{amsmath,bm,natbib,xcolor,mathtools}
\usepackage[colorlinks=true,citecolor=blue,urlcolor=blue]{hyperref}

\begin{document}	
	\title{Chain Stretch Relaxation from Low Frequency Fourier Transform Rheology}
	
	\author{C. D. Reynolds}
	\affiliation{Department of Chemistry, Durham University, Durham, DH1 3LE, United Kingdom}
	\affiliation{Department of Chemistry, The University of Birmingham, Edgbaston, Birmingham, B15 2TT, United Kingdom}
	\author{D. M. Hoyle}
	\affiliation{Department of Chemistry, Durham University, Durham, DH1 3LE, United Kingdom}
	\affiliation{Offshore Renewable Energy Catapult, Blyth, Northumberland, NE24 1LZ}
	\author{T. C. B. McLeish}
	\affiliation{Department of Physics, Durham University, Durham, DH1 3LE, United Kingdom}
	\affiliation{Department of Physics, York, United Kingdom}
	\author{R. L. Thompson}
	\affiliation{Department of Chemistry, Durham University, Durham, DH1 3LE, United Kingdom}
	
	\date{\today}
	
	\begin{abstract}
		
		Medium or large amplitude oscillatory shear (M/LAOS) is sensitive to polymer chain structure, yet poses unsolved challenges for {\em a priori} structural characterisation. We present a MAOS protocol applied to near-monodisperse linear polymer melts, from which chain-stretch relaxation, a key structural feature, is discernible.  The third harmonics of MAOS frequency sweeps are decomposed into real and imaginary components and found to obey time-temperature superposition. Significantly, these third harmonic features occur at low frequency and are readily accessible with standard rheometers. For materials where phase transitions restrict the use of time temperature superposition, this method has potential to greatly increase the scope of rotational rheometry for structural analysis of polymers. However, the relationship between MAOS data and characteristic relaxation times is complex and to elucidate this, a modelling approach is required. The GLaMM molecular tube-based model of linear entangled melt rheology and structure, which has no free parameters, closely follows the form of our experimental results for the third harmonics and contains discriminatory features which depend only on the polymer's chain stretch relaxation time. However, we find fundamental differences in magnitude and the frequency dependence of the third harmonics which must be resolved in order to fully understand the molecular basis of the stress response, and quantitatively study chain stretch. 
		
	\end{abstract}
	
	
	\maketitle
	
	{\em Introduction:  } 
	Oscillatory shear rheology is sensitive to the microstructure of complex soft materials (e.g. polymers \cite{Malmberg2002,Vega1998,Vega1999} or immiscible blends \cite{Zhang2008a}). The technique subjects a fluid sample to oscillatory shear strain at a given amplitude and frequency, $\gamma(t) = \gamma_0\sin(\omega t)$ and analyses the stress response. Medium or large amplitude oscillatory shear (collectively henceforth LAOS) is defined as having $\gamma_0$ sufficient to induce detectable levels of non-linear stress response (for a review see \citet{Hyun2011}). 
	
	LAOS reveals microstructural information (additional to that available from linear rheology) in a range of viscoelastic materials, e.g. gels and networks \cite{Ng2011,Martinetti2014,Bharadwaj2017},
	wormlike micelles \cite{Dimitriou2012}, soft glasses \cite{Renou2010}, emulsions \cite{Duvarci2017}, particle suspensions \cite{Li2009} and biological fluids \cite{Ewoldt2008}. 
	For polymers in particular, LAOS is complementary to small-amplitude measurements and results are a complex function of molecular architecture such as linear \cite{Cziep2016}, star  \cite{Song2016} and comb architectures \cite{Wagner2011} and can be used to quantify the level of branching in industrial resins such as metallocene-catalysed sparsely branched HDPEs \cite{Hoyle2014,Vittorias2006} or tubular-reacted randomly branched LDPEs \cite{Vittorias2006,Abbasi2013}. However and crucially, LAOS is not yet a standard analytical tool for characterising molecular architecture.  
	
	The stress response of a complex material in LAOS can be characterised in several ways. The transient shear stress can be plotted against time and visual distortions from a sinusoidal curve can be seen \cite{Clemeur2003}. However, the shear stress is more commonly plotted against strain to give Lissajous-Bowditch curves \cite{Giacomin1993a,Ewoldt2010} and these are grouped together for various frequencies and strain-amplitudes in Pipkin diagrams \cite{Ewoldt2008} which give a visual ``fingerprint" of a material. For more quantitative analysis, the stress is typically decomposed into some series such as Fourier \cite{Hyun2011}, Chebyshev polynomials \cite{Ewoldt2008} or graphical interpretations \cite{Cho2005}. It's worth noting that no dominant methodology is established as a benchmark analysis method.
	In this letter we focus on Fourier transform rheology (FTR) as this has been shown to be a sensitive enough technique to isolate small non-linearities in the material stress response, either from shear stress \cite{MacSporran1984,Wilhelm1998,Wilhelm1999,Wilhelm2002,Hyun2011} or the first normal stress \cite{Nam2008}.
	
	In FTR the shear stress response to the imposed sinusoidal shear-rate is expressed as a Fourier series, $\sigma_{xy}^{FT} = \sum_{n}\left[I'_n\sin(n\omega t)+I''_n\cos(n\omega t)\right]$ where $I'_n$ and $I''_n$ are the Fourier coefficients. The complex dynamic moduli for the $n^{th}$ harmonic are defined as $G'_n=I'_n/\gamma_0$ and $G''_n=I''_n/\gamma_0$. For $n=1$ we recover the standard storage and loss moduli: $G'=I'_1/\gamma_0$ and $G''=I''_1/\gamma_0$ as $\gamma_0 \rightarrow 0$. 
	Non-linearities in LAOS are measured typically through the third harmonics ($n=3$) since by symmetry the even harmonics are zero \cite{Singh2018} (we use the $2nd$ harmonic to measure the noise).
	Popular reported quantities are the absolute value of the third harmonic $I_3 = \sqrt{I'_3 + I''_3}$ and its value as a ratio to the absolute first harmonic, $I_{3/1}=I_3/I_1$, which is often plotted as a function of increasing $\gamma_0$ \cite{Wilhelm2002}. 
	Secondarily considered, is the phase shift of the third harmonic $\Phi_3 = \phi_3 - 3\phi_1$, with $\tan(\phi_n)=\frac{G''_n}{G'_n}$ \cite{Vittorias2006} and $Q = I_{3/1}\gamma_0^{-2}$ which plateaus to a constant value ($Q_0$) in the limit of small amplitude \cite{Hyun2008}. $Q_0$ can also be decomposed into $Q'_0$ and $Q''_0$ \cite{Song2018}. 
	Non-linearities in the first harmonic have been used to characterise behaviour at high strains \cite{Song2019} and to predict the form of the third harmonic \cite{Carey-DeLaTorre2018}. However, we choose to focus on the third harmonic where non-linearities occur at small strains and are experimentally accessible.
	Of recent interest is the MAOS protocol \cite[e.g.][]{Singh2018}, defined as the strain regime within which the imposed $\gamma_0$ is sufficient for non-linearities in the stress to be experimentally measured yet maintain the relation $I_{3/1} \propto \gamma_0^2$. 
	
	Interpretation of LAOS-FTR results relies on comparison with some relevant constitutive theory, since this method does not explicitly reveal a direct relationship between material microstructure and the subsequent higher harmonic dependency \cite[several are covered in][]{Hyun2011}.
	However, constitutive modelling of LAOS is comparatively underdeveloped. Hyun\cite{Hyun2007} compared several constitutive models such as Giesekus, Phan-Tien Tanner and the Pom-pom model. An example using the Giesekus model is given in the Appendix. The model contains a non-linearity factor ($\alpha$) which can be fit to the MAOS response, but cannot capture all flows (transient shear, extension and MAOS) with a single value. The Pom-pom model has been used to characterise branching \cite{Vittorias2006,Clemeur2003,Hoyle2014} which has been effective due to its structure based construction and parametrisation. The molecular stress-function (MSF) theory has also been shown to be capable of capturing extensional and LAOS rheology simultaneously \cite{Abbasi2013}.
	For the MAOS regime both the Pom-pom and MSF theories have been shown to broadly capture the intensity of $I_{3/1}$ over a range of frequencies for a range of materials from monodisperse linear, star-arm to randomly branched polymers \cite{Cziep2016,Song2016,Hoyle2014}.
	A limitation of all these approaches is that fitting is required to match theory to LAOS data; typically the parameters are set by a different rheometric experiment and the fits are ``multi-modal'' in form. 
	These factors obscure the true molecular response and hence the ability of the LAOS technique to inform on structure. 
	
	The aim of this letter is to  
	(i) present the rich phase information that is contained in the third harmonic and show that this can be meaningfully used to characterise the molecular rheology of well-defined materials (including parameters that can be extremely difficult to obtain from linear rheology), and
	(ii) compare these new results and a multi-modal modelling approach with the most detailed truly molecular constitutive model currently available.
	
	
	{\em Materials and Experimental:  } Linear polybutadiene chains of defined molecular weights were synthesised by standard living anionic polymerisation \cite{Mykhalyk2011}. In table \ref{table::materials} we detail the material parameters measured by Gel Permeation Chromatography (GPC) and standard oscillatory rheology. 
	\begin{table}[ht]
		\centering	
		\caption{Material parameters for the polyisoprene samples at a reference temperature of $25^{\circ}$C.}
		\begin{tabular*}{0.48\textwidth}{@{\extracolsep{\fill}} c | c c c c c}
			\hline \hline
			Sample name & $Mw$ [kg/mol] & PDI [-] & $\eta^*_0 [Pa.s]$ & $\tau_d [s]$ & $Z$ \\
			\hline
			PI20k  & 21.5 & 1.02 & 126 & 0.00058 & 5 \\
			PI100k & 100  & 1.03 & 31,600 & 0.155 & 21 \\
			PI150k & 145  & 1.02 & 113,000 & 0.55 & 30 \\
			PI400k & 387  & 1.05 & 2,910,000 & 13.8 & 80 \\
			\hline \hline
		\end{tabular*}
		\label{table::materials}  
	\end{table}
	
	Rheological experiments were performed on a Discovery HR-2 (TA Instruments) equipped with an environmental test chamber supplied with liquid nitrogen. For linear rheology, a 25 mm parallel plate geometry was used and the dynamic moduli were measured using frequency sweeps ($10^{-2}\textrm{Hz} \le \omega \le 10^2\textrm{Hz}$ and $1\% \le \gamma_0 \le 5\%$) at various temperatures between $-30^{\circ}$C and $50^{\circ}$C. The results at each temperature were shifted to a reference temperature of $25^{\circ}$C using WLF theory \cite{Ferry1980} and RepTate software \footnote{\url{http://www.reptate.com}}. 
	MAOS measurements were carefully made in seperate experiments using a 25 mm, $4^{\circ}$ cone. Frequency sweeps were  performed for strain amplitudes of $5\% \le \gamma_0 \le 20\%$ and Frequencies under $5$ rad/s to limit inertial and instrument effects. Transient data was recorded and the Fourier coefficients for the stress were extracted from the transient stress data using an  in-house MATLAB program \footnote{\url{https://sourceforge.net/projects/cdrheo/}} which uses a Fast Fourier Transform routine. 
	Care was taken during sample preparation and measuring to ensure the accuracy of the LAOS results, with details of these protocols given in the SI and in \cite{Reynolds2018a}.
	
	{\em Modelling:  } 
	We compare the polyisoprene rheology to a self-consistent set of constitutive equations that transition from linear to non-linear theory using the the same underlying concepts. Firstly, the Likhtman-McLeish linear theory \cite{Likhtman2002} accurately describes the full relaxation pathways of linear polymer chains subjected to a linear deformation. This theory is then extended to the GLaMM model which considers non-linear stress response of the whole chain using a series of well-considered approximations and closure assumptions. The GLaMM \cite{Graham2003,Auhl2008a} model offers a sophisticated treatment of mono-disperse linear polymer melts that includes several relaxation mechanisms; chain diffusion, chain stretch, convective constraint release and contour length fluctuations. Although the GLaMM model captures non-linear rheology without need for fitting free parameters, it is computationally expensive and for simulations with any spatial variance in the flow rates, it is more convenient to use the coarse-grained version of the model: the Rolie-Poly \cite{Likhtman2003} model. This considers the chain as a single end-to-end vector, as opposed to the contour dependence included in the GLaMM model. We consider both a one-mode model and a multi-mode model. The multimode version of the Rolie-Poly model is used to restore the transient features lost in removing higher frequency chain-motion during coarse graining of the GLaMM model. However, the multimode Rolie-Poly model must be fitted either against GLaMM predictions or experimental data for rheological linear and non-linear flows.
	
\begin{figure}[t]
		\includegraphics[width=0.48\textwidth ]{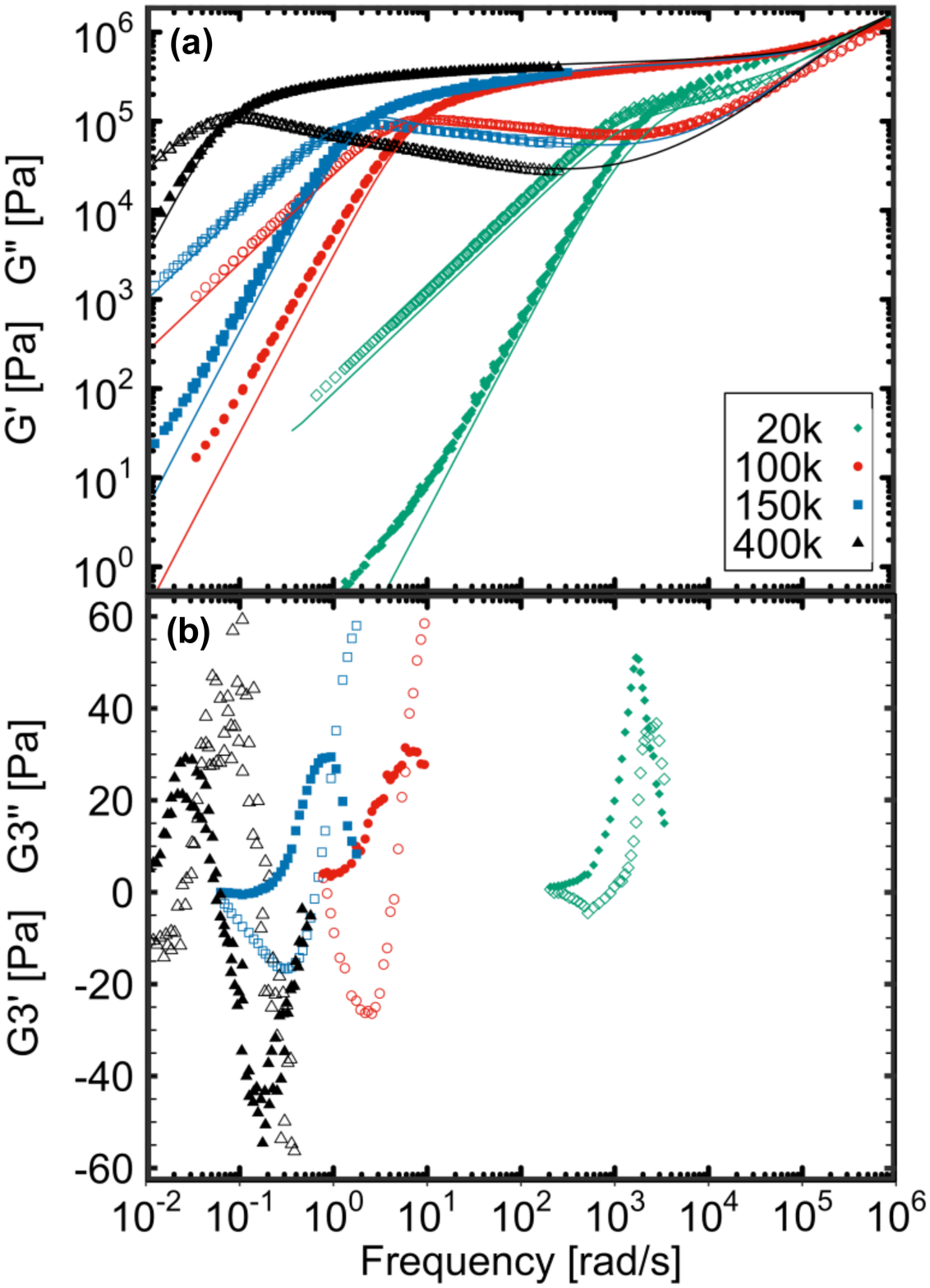}
		\caption{(a): linear rheology for the four PI blends at $25^{\circ}$C detailed in table \ref{table::materials} with comparison to Likhtman-McLeish linear theory. (b): MAOS rheology for the four blends at $25^{\circ}$C with a strain of $20\%$ where the open symbols are $G'_3$ and the closed = $G''_3$'. For PI400k the results are a TTS for temperatures in the range of $25^{\circ}$C-$50^{\circ}$C. \label{fig::1}}
	\end{figure}

	{\em Results:  } 
	We consider the linear and non-linear oscillatory shear for four molecular weights of polyisoprene. All the samples are entangled; with one sample weakly entangled (number of entanglements per chain, $Z=5$), two samples moderately entangled ($Z=21$ and $Z=30$) and one sample highly entangled ($Z=80$).
	Figure \ref{fig::1} shows the linear rheology of the four materials at $25^{\circ}$C with properties given in table \ref{table::materials}. For all materials the terminal cross-over  was experimentally accessible giving the reptation relaxation time $\tau_d$ and  for PI20K and PI100k the high frequency cross-over was also measured, giving the entanglement relaxation time $\tau_e$. The only characteristic difference between the samples is that the weakly entangled PI20k shows no minimum in $G''$. Also in this figure are the predictions of the Likhtman-McLeish linear theory, without adjustable parameters (lines) which gives an excellent prediction of the linear rheology for all four samples with the only parameter differentiating them being the molecular weight. 
	
	In the bottom half of figure \ref{fig::1} the real and imaginary parts of the third harmonic are plotted for each PI sample. PI20k-PI150k were measured at $25^{\circ}$C with a strain of $20\%$ and PI400k at higher temperatures (up to $50^{\circ}$C) transposed to $25^{\circ}$C with time-temperature superpostion. All sets of curves follow similar qualitative behaviour, with shifts in frequency with molecular weight in a manner that tracks the linear rheology.  The key feature of the third harmonic curves is a cross-over in $G'_3$ and $G''_3$, which is always found at a lower frequency than the cross-over associated with the terminal relaxation time in the linear rheology (and orders of magnitude lower than the frequency equal to the inverse chain stretch time).  For the moderately and highly entangled polymers, the cross-over occurs close to the peak value of G3' and the inflection point of G3''. However, the weakly entangled PI20k has a different shape, with $G'_3$ rising to a higher maxima before falling to cross-over at the lower peak value of $G''_3$.  A plot of figure \ref{fig::1} with frequency normalised by $\tau_d$ and details of the TTS of PI400k is given in the Appendix.

	{\em Discussion:  } 
	The LAOS results in figure \ref{fig::1} show several non-linear features including a cross-over between $G'_3$ and $G''_3$, and extrema in both. 
	These features move to lower frequencies with increasing $Mw$. Plotting the results against Deborah number $De = \omega\tau_d$ collapses the terminal linear rheology but not the LAOS features [c.f. Appendix, figure 2]. Therefore, LAOS features are associated with faster relaxation processes. The cross-over in $G'_3$ and $G''_3$ moves to lower $De$ with increasing $Mw$ and further occurs for $De < 1$, hence at lower frequencies than the characteristic reptation rate. 
	All the theories reported here predict the 3rd harmonic cross-over to occur at $De > 1$.
	
	To understand these complex rheological structures comparison with a non-linear viscoelastic theory, which simultaneously describes the linear rheology is essential. This will (i) test our current understanding of the underlying polymer physics and (ii) allow these measurements to be used as an analytical technique.

\begin{figure}[t]
		\includegraphics[width=0.48\textwidth ]{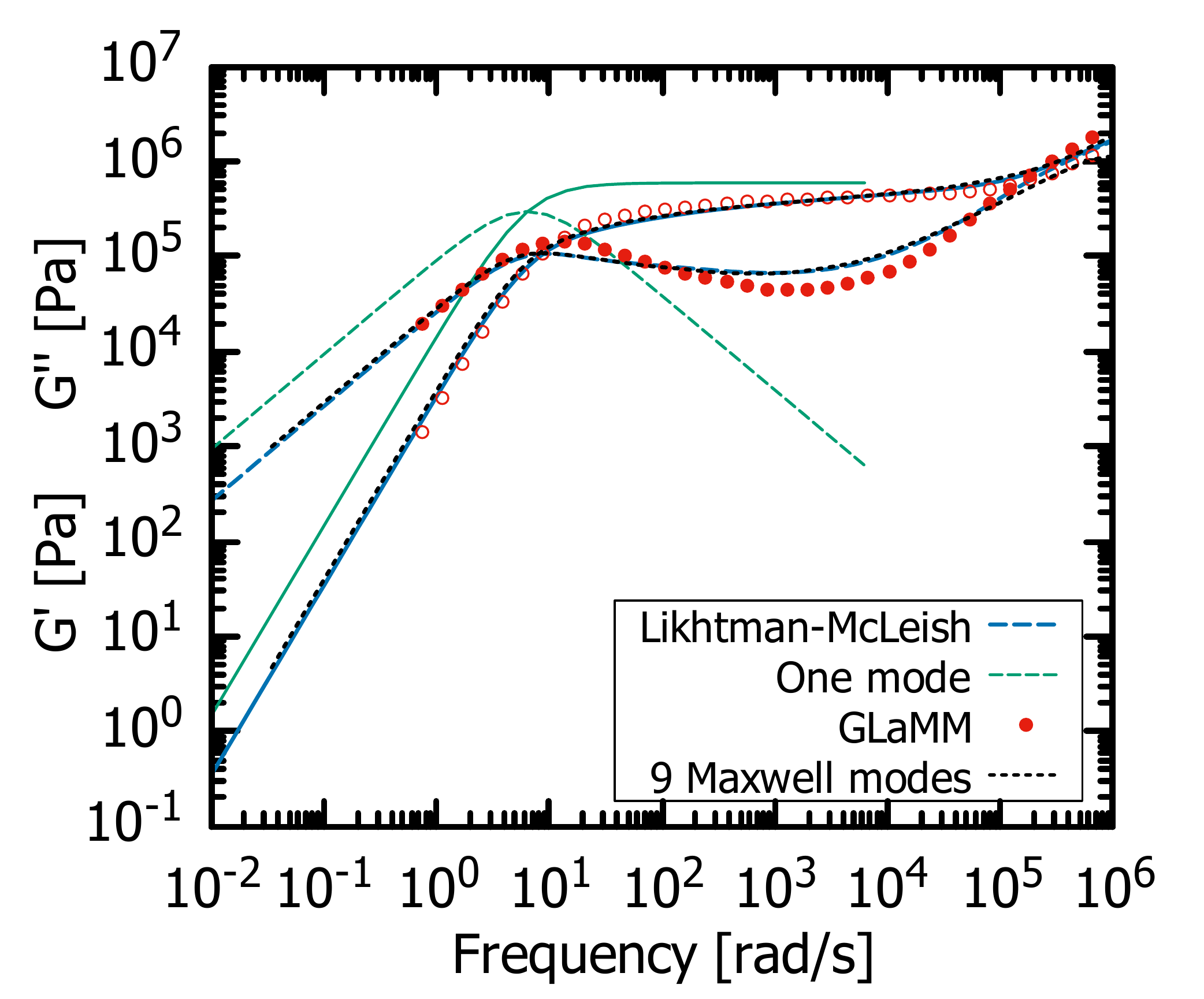}
		\caption{Linear dynamic moduli predictions for PI100K at $25^{\circ}$C, using Likhtman-McLeish Theory, GLaMM and RoliePoly with 1 and 9 Maxwell modes. Details of the parameters used are provided in the Appendix.  \label{fig::2}}
		\includegraphics[width=0.48\textwidth ]{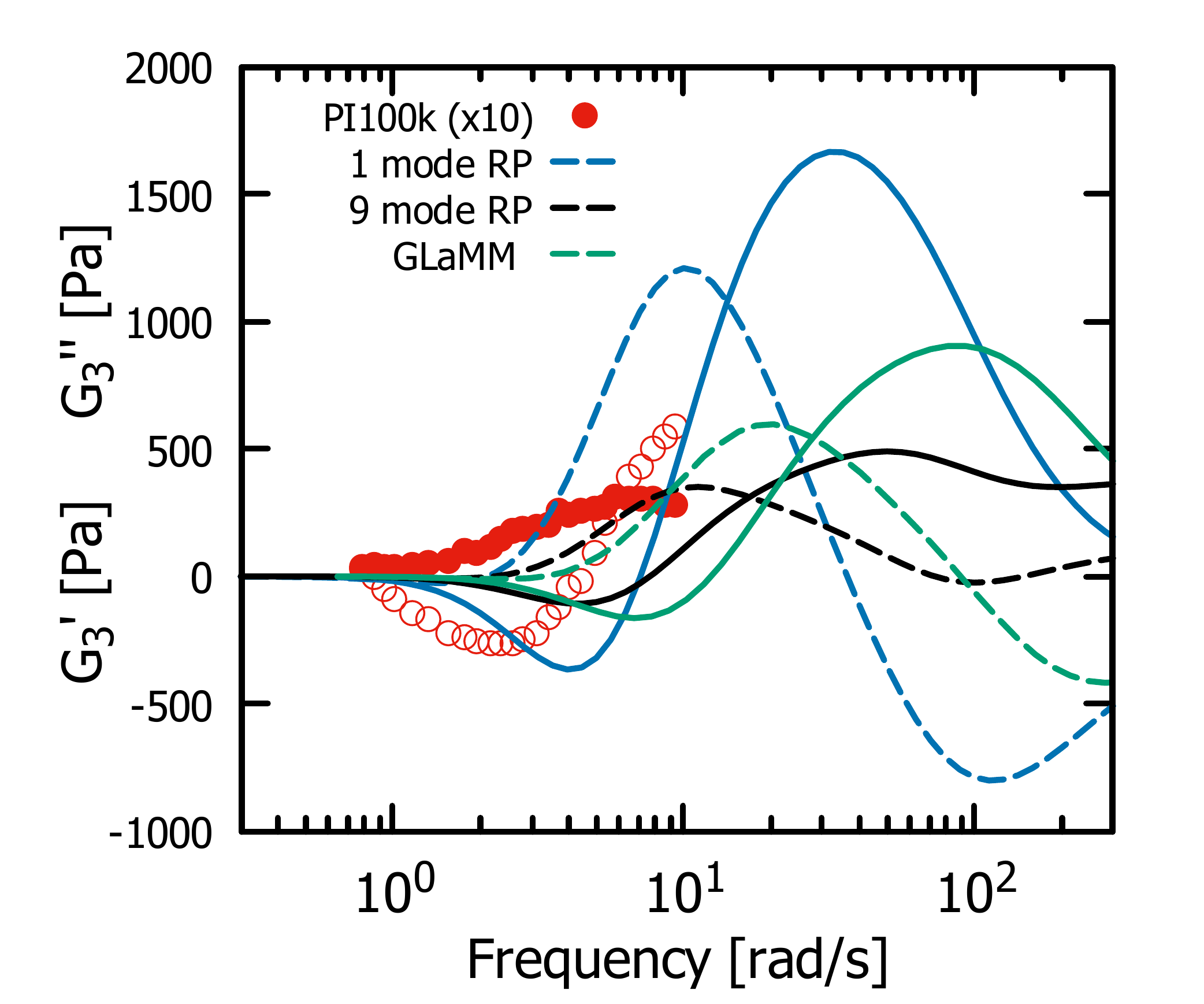}
		\caption{GLaMM predictions of third harmonics compared to RoliePoly with 1 and 9 modes and experimental results. The open symbols and solid lines are $G'_3$ and the filled symbols and dashed lines are for $G''_3$'. \label{fig::3}}
	\end{figure}

	We now compare the previously introduced molecular rheology theories to the experimental results for the PI100k sample at $25^{\circ}$C, figure \ref{fig::2}. We consider the GLaMM model, a one-mode Rolie-Poly (1-RP) model and a multi-mode Rolie-Poly (9-RP) model with 9 Maxwell modes. 
	Comparing the linear rheology predicted by Likhtman-McLeish (LM) linear theory which nearly superimposes onto the experimental data in figure \ref{fig::1}, we see the non-linear extension of the LM model follows both moduli closely with a slight deviation in $G''$ near the cross-over point. The one-mode RP model is calculated with the moduli and reptation time taken from LM theory and hence captures little of the linear rheology except the correct scaling in the terminal region. Finally, the 9-RP model (fitted to experimental data) closely echoes the data over the complete frequency range until the final mode at around $10^5$ [rad/s]. 
	
\begin{figure}[t]
		\includegraphics[width=0.48\textwidth ]{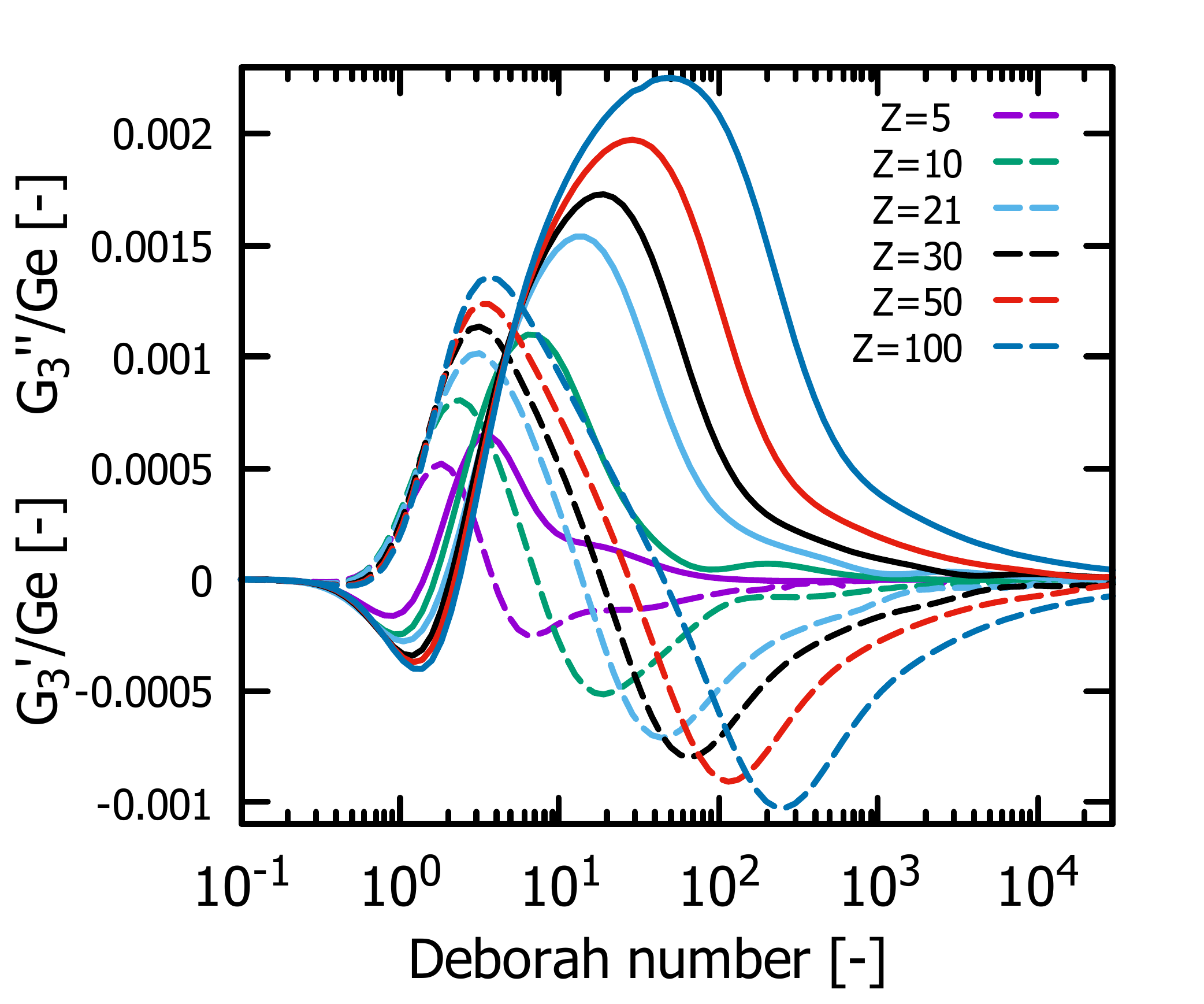}
		\caption{GLaMM predictions of third harmonics with varied entanglement number. The solid lines are $G'_3$ and the dashed lines are for $G''_3$'. \label{fig::4}}
	\end{figure}

	In figure \ref{fig::3} we compare the non-linear theories to the LAOS data. We multiplied the modulus of the experimental data by a factor $10$ for it be discernable on the plot. While all theories reproduce the qualitative shape of $G_3'$ and $G_3''$, it is clear that all theories overestimate the amplitude of the third harmonic by at least a factor 10. It is also clear that the cross-over between $G3'$ and $G3''$ occurs at a significantly higher frequency for all theories compared to experiment. We have checked various parameters such as the convective constraint release (CCR) rate, order one parameter ($R_s$) and the GLaMM discretisation ($N$) which have minor effects on the magnitude of the third harmonics but show no qualitative differences to those presented here (c.f. Appendix). Changing the parameter $R_s$ has the effect of changing the stretch relaxation whilst preserving the linear rheology. Even changing this parameter (effectually reproducing figure \ref{fig::4}) cannot bring the $3$rd harmonic cross-over below a frequency less than $\tau_d^{-1}$.
	
	We can see in figure \ref{fig::4} that there are clear trends in the features of the third harmonics that are a function of molecular weight (or $Z$). The theory predicts that with greater Z, we see an increase in the magnitude of both the peaks in $G'_3$ and $G''_3$ and the negative minima in $G''_3$, as well as a shift of all features to higher $De$. This is counter to the experimental observations which show the features moving to lower $De$ with molecular weight (seen in  Appendix figure 2) and decreasing or remaining as constant magnitude. It is clear however that the theories do capture the correct form of the experimental data, capturing all of the features noted in experiment.
	
	Although the single mode Rolie-Poly model captures little of the linear rheology, it reproduces all the features in the MAOS sweeps, albeit with higher magnitudes. The simplicity of this model means we can extract analytical forms for the real and imaginary components of the third harmonic in the limit of small strain amplitude (given in the Appendix). Although complex, these expressions are only dependent on the oscillation frequency and the Rouse time of the polymer. This implies that the behaviour we see, qualitatively captured by this simple model, is driven by chain stretch. This is significant as the frequencies used are well below the inverse Rouse time of the polymers, and so no contribution from Rouse behaviour is expected in linear rheology. Moreover, the Rouse behaviour of these polymers is difficult to access at all in linear rheological tests (Here it required temperatures of $-30^{\circ}$C). MAOS therefore is a tool for probing Rouse behaviour at low frequency, which is beneficial for systems where high frequencies are inaccessible, or for semi-crystalline polymers, where TTS is restricted to temperatures above the melting point.
	
	{\em Conclusions:  } 
	We report a MAOS protocol, alongside evidence that the behaviour of polymer melts in MAOS is driven by chain stretch. This method makes polymer chain stretch behaviour accessible at low frequencies on standard torsional rheometers. 
	The standard dynamic moduli for well entangled polymers ($Z\geq10$) superimpose for Deborah numbers $De \leq 10$. The MAOS results show that the samples can be differentiated by probing their weak non-linear response at $De \lesssim 1$. 
	%
	The MAOS results can differentiate these samples from their non-linear response revealing more characteristic properties than linear rheology alone.  
	The key characteristic flow time scales are the orientation and stretch time ($\tau_d$ and $\tau_r$ respectively) and these are related to the entanglement time and the number of entanglements. The terminal relaxation cross-over allows $\tau_d$ to be easily obtained (along with $G_e$) and determining any of $Z$, $\tau_e$ or $\tau_r$ instantly gives the full rheological map of flow properties.
	
	The MAOS measurements clearly access the faster non-linear relaxation mechanisms, usually measured at $De\gg1$, at Deborah numbers O(1). 
	An example of the potential of this approach is for semi-crystalline polymers where the temperature range accessible in melt rheology is very limited. This restriction is heightened by the often large difference between the glass-transition temperature and the melt-transition temperature which reduces the effect of temperature change on viscosity and therefore severely limits the effects of time-temperature superposition theories. However, a modelling approach is required for quantitative study of this behaviour, in order to fully establish the complex relationship between this response and the characteristic relation times.
	
	We have shown a significant discrepancy in the predictions of MAOS with Rolie-Poly and GlaMM models, the latter notable for having no adjustable parameters. Qualitatively, the behaviour in MAOS is captured by both models, yet remarkably neither quantitatively matches the experimental data for these simple model materials. This indicates an addition is required, even to the current gold standard in rheological models to fully capture MAOS data.
	
	{\em Acknowledgements:  } The authors gratefully acknowledge financial support from Michelin R\&D (Material Performance and Processability) and M. Oti for providing the materials.

\clearpage
\onecolumngrid
{\large \bf Appendix: Chain Stretch Relaxation from Low Frequency Fourier Transform Rheology}
\vspace{0.5cm}
\makeatletter
\renewcommand{\bibnumfmt}[1]{[A#1]}
\renewcommand{\citenumfont}[1]{A#1}

{\em Details on sample preparation and measurement protocols: } 
Samples were prepared for rheometry by pressing into discs $1$ mm thick and $25$ mm in diameter under a force of 4 tonnes at room temperature followed by equilibration for at least 10 minutes. The disc was loaded into the rheometer and the geometry driven to the sample gap + $5\%$. The normal force was left to dissipate to $\le 0.1$N and the sample trimmed after which the geometry was driven to the final sample gap and the normal force again allowed to dissipate before measurement.

Care was taken to ensure the accuracy of the LAOS measurements: (i) many cycles were averaged over (typically 100 plus 5 startup cycles which were discarded) to minimise the noise in the system (measured using the 2nd harmonic which should be zero by symmetry \cite{Wilhelm2002-SI}), (ii) the effects of edge fracture and slip were avoided by monitoring the sample and the magnitudes of the even harmonics and comparison of $G'$ and $G''$ with the linear rheology and (iii) superharmonic superposition \cite{Poulos2015-SI} was measured for the instrument and avoided by performing measurements at frequencies under $5$ rad/s.

{\em Details on constitutive models used: } 
The Likhtman-McLeish linear model \cite{Likhtman2002-SI} was compared to the linear rheology of the PI samples in the main paper. This was not solved directly, rather using Reptate software. The following equation was used from the LM model to relate the entanglement time $\tau_e$,
\begin{equation}
\tau_d = 3Z^3 \left( 1 - 2\frac{C_1}{\sqrt{Z}} + \frac{C_2}{Z} - \frac{C_3}{Z^{3/2}} \right) \tau_e.
\label{eqn:td}
\end{equation} 
The coefficients are $C_1 = 1.69$, $C_2 = 4.17$, and $C_3 = -1.55$.

We give and overview of the GLaMM model with full details found in \cite{Graham2003-SI,Auhl2008a-SI}. We calculate the stress tensor as follows,
\begin{equation}
\bm\sigma = \frac{12G_e}{5Z}\int_{0}^{Z}\bm f(s,s')ds + \frac{G_e}{Z}\int_{-\infty}^{t}\sum_{p=Z}^{N} \exp\left(-\frac{2p^2(t-t')}{Z^2\tau_e}\right)\left[\bm K(t') + \bm K(t')^T\right]dt',
\end{equation}
where the first term in the R.H.S is the stress relaxation at length-scales larger than an entanglement and the second term is the entanglement Rouse relaxation with relxation time $\tau_e$. $G_e$ is the modulus, $Z$ is the number of entanglements, $\mathbf{K}$ is the velocity gradient tensor and $p$ is the Rouse mode number. The $\bm f(s,s')$ tensor is a tangent correlation function,
\begin{equation}
f_{\alpha\beta}(s,s';t) \equiv  \left\langle \frac{\partial R_{\alpha}(s,t)}{\partial s}\frac{\partial R_{\beta}(s',t)}{\partial s'} \right\rangle,
\end{equation}
which considers the conformation of the polymer chain described by vector $\mathbf{R}(s,t)$ between points $s$ and $s'$ with $\{s,s'\}\subset[0,Z]$. The evolution of $\bm f(s,s')$ is given by,
\tiny
\begin{equation}
\begin{aligned}
\frac{\partial}{\partial t}f_{\alpha\beta}(s,s';t) = 
\left(\kappa_{\alpha\gamma}f_{\gamma\beta} + f_{\alpha\gamma}\kappa_{\gamma\beta} \right)
+  \frac{1}{3\pi^2Z\tau_e}\left(\frac{Z}{Z^*(t)}\right)^2
\left(\frac{\partial}{\partial s}+\frac{\partial}{\partial s'}\right)
\frac{D^*(s,s')}{\lambda(s,s')}
\left(\frac{\partial}{\partial s}+\frac{\partial}{\partial s'}\right)f_{\alpha\beta} + \cdots \\
\cdots + \frac{3a\nu}{2}
\left[\frac{\partial}{\partial s}\frac{1}{\lambda(s)}\frac{\partial}{\partial s}
(f_{\alpha\beta} - f^{eq}_{\alpha\beta}) +
\frac{\partial}{\partial s'}\frac{1}{\lambda(s')}\frac{\partial}{\partial s'}
(f_{\alpha\beta} - f^{eq}_{\alpha\beta}) \right]
+ \frac{R_s}{2\pi^2\tau_e}\left[\frac{\partial}{\partial s}\left(f_{\alpha\beta}\frac{\partial}{\partial s}\ln \lambda^2(s) \right) +
\frac{\partial}{\partial s'}\left(f_{\alpha\beta}\frac{\partial}{\partial s'}\ln \lambda^2(s') \right) \right].\\
\end{aligned}
\label{eqn::f_func}
\end{equation}
\normalsize
Equation \ref{eqn::f_func} contains four terms which in order are convection, reptation and contour length fluctuations, constraint release and retraction. The equation contains parameters for diffusion ($D*(s,s')$), and a retraction rate ($\lambda(s)$), with constants for CCR ($\nu$) and $R_s$, a geometric parameter of order unity. By considering the inter-segmental motions of the chain in Fourier space, the GLaMM equation can be simplified by considering only the first Fourier mode. This removes the $s$ dependency and the chain becomes dumbbell like. The resulting model is the Rolie-Poly model which is given by,
\begin{equation}
\frac{d\bm{\sigma}}{dt} = \bm{K}\cdot\bm{\sigma} + \bm{\sigma}\cdot\bm{K}^T - \frac{1}{\tau_d} \left( \bm{\sigma} - \bm{I}\right) - \frac{2\left(1-\sqrt{3/tr\bm\sigma}\right)}{\tau_r}\left(\bm\sigma + \beta\left(\frac{tr\bm\sigma}{3}\right)^\delta(\bm\sigma - \bm I)\right).
\label{eqn::RP}
\end{equation}

To recover the transient stress growth in both non-linear shear and extensional flow tests we use the multimode Rolie-Poly model. The faster modes are used to recover the correct viscoelastic envelope [c.f. figures \ref{fig::8} and \ref{fig::9}] and the non-stretch limit of the RoliePoly equation is used for these 'fast' modes (where stretch is relaxed essentially infinitely fast);
\begin{equation}
\frac{d\bm{\sigma}}{dt} = \bm{K}\cdot\bm{\sigma} + \bm{\sigma}\cdot\bm{K}^T - \frac{1}{\tau_d} \left( \bm{\sigma} - \bm{I}\right) - \frac{2}{3}tr\left(\bm K \cdot \bm\sigma\right)\left(\bm\sigma + \beta(\bm\sigma - \bm I)\right).
\end{equation}

The simplicity of the Rolie-Poly model (eqnn. \ref{eqn::RP}) allows some analytic progress to be made.
If we expand the (dimensionless) extra stress tensor in increasing powers of strain amplitude we can derive an expression of the Fourier coefficients $I^{',''}_n$ that are applicable in the MAOS regime of flow. More details can be found in \cite{Hoyle2010a-SI,Hoyle2014-SI}, where a similar approach was performed on the Pom-pom equations. Formulae for the third harmonics of the Rolie-Poly in limit of small strain amplitude with $\beta = 0$ (and $\tau_d = 1$ for clarity such that each component is dimensionless) are given by:
\tiny
\begin{equation}
I_3'  = -\frac{1}{6}\frac
{(36\tau_r\omega^6 - 73\tau_r\omega^4 - 39\omega^4 - 35\tau_r\omega^2 - 13\omega^2 + 2\tau_r + 2)\omega^3}
{144\omega^{10}\tau_r^2 + 376\omega^8\tau_r^2 + 72\tau_r\omega^8 + 36\omega^8 + 377\tau_r^2\omega^6 + 
	170\tau_r\omega^6 + 85\omega^6 + 123\tau_r2\omega^4 + 126\tau_r\omega^4 + 63\omega^4 + 19\tau_r^2\omega^2 + 30\tau_r\omega^2 + 15\omega^2 + \tau^2 + 2\tau_r + 1},
\label{eqn::I3'}
\end{equation}
\normalsize
and 
\tiny
\begin{equation}
I_3'' =  \frac{1}{6}\frac
{(96\tau_r\omega^4 + 8\tau_r\omega^2 + 18\omega^4 - 17\omega^2 - 15\tau_r - 11)\omega^4}
{144\omega^{10}\tau_r^2 + 376\omega^8\tau_r^2 + 72\tau_r\omega^8 + 36\omega^8 + 377\tau_r^2\omega^6 + 170\tau_r\omega^6 + 85\omega^6 + 123\tau_r2\omega^4 + 126\tau_r\omega^4 + 63\omega^4 + 19\tau_r^2\omega^2 + 30\tau_r\omega^2 + 15\omega^2 + \tau^2 + 2\tau_r + 1}.
\label{eqn::I3''}
\end{equation} 
\normalsize
Indeed, these equations (\ref{eqn::I3'} and \ref{eqn::I3''}) give the same result as seen in figure 3 for the one-mode RoliePoly, i.e. qualitatively similar to the data but quantitatively different in amplitude and frequency dependence. 
Clearly, the relationship between chain stretch relaxation time and frequency is complicated and which would explain the previous difficultly in extracting molecular information from LAOS.

\begin{figure}[ht]
	\includegraphics[width=0.6\textwidth ]{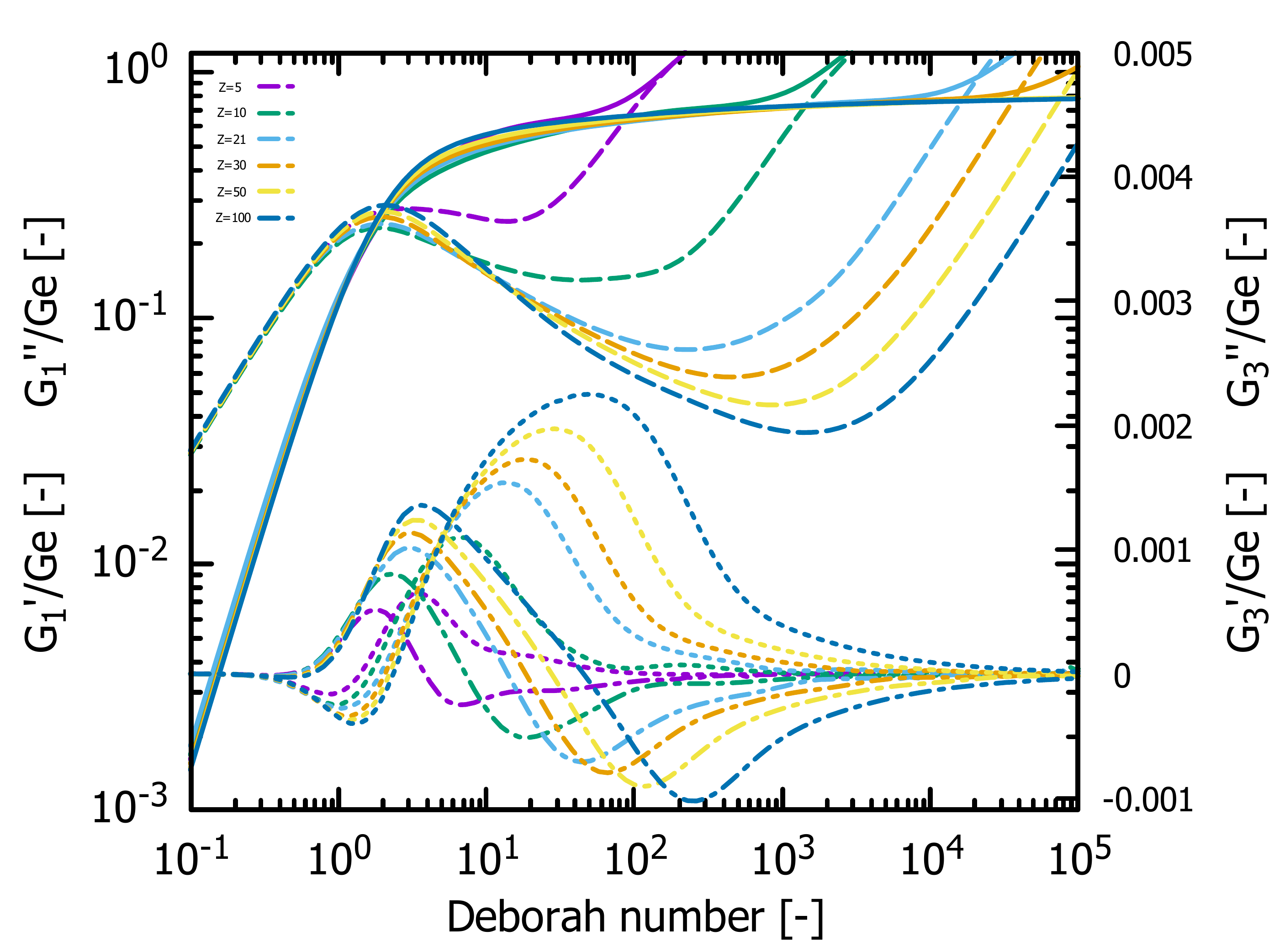}
	\caption{Counterpart to figure 4 in the main article. Variations in $G3`$ and $G3``$ as a function of $Z$ with comparison to the dynamic moduli. Of particular note is the characteristic cross-over frequencies for each set of harmonics. In contrast to the experimental data the cross-over for the third harmonics occur at a higher frequency than the standard cross-over. \label{fig::5}}
\end{figure}

\begin{table}[ht] 
	\caption{Likhtman-McLeish theory parameters for polyisoprene at $25^{\circ}C$ that are common to all molecular weights.  \label{tab::LMth}}
	\begin{tabular}{ @{\extracolsep{\fill}} c c c c }
		\hline 
		\phantom{aaaaa}$G_e$\phantom{aaaaa} & \phantom{aaaa}$M_e$\phantom{aaaa} & \phantom{aaaa}$\tau_e$\phantom{aaaa} & \phantom{aaaa}$c_\nu$\phantom{aaaa}  \\ 
		\hline 
		\hline
		5.9558E5  & 4.8158 & 1.321E-5 & 0.1  \\ 
		\hline
	\end{tabular} 
\end{table}
\begin{table}[ht] 
	\caption{GLaMM theory parameters for simulations.  \label{tab::GLaMM}}
	\begin{tabular}{ @{\extracolsep{\fill}} c c c c }
		\hline 
		\phantom{aaaaa}$G_e$\phantom{aaaaa} & \phantom{aaaa}$\tau_d$\phantom{aaaa} & \phantom{aaaa}$R_s$\phantom{aaaa} & \phantom{aaaa}$c_\nu$\phantom{aaaa}  \\ 
		\hline 
		\hline
		1.0  & 1.0 & 2.0 & 0.1  \\ 
		\hline
	\end{tabular} 
\end{table}
The parameters used for Likhtman-McLeish and GLaMM theories are shown in tables \ref{tab::LMth} and \ref{tab::GLaMM}, respectively. The other key parameter for the GLaMM simulations is the entanglement number $Z = \dfrac{M_w}{M_e}$ from which the entanglement time $\tau_e$ is derived from equation \ref{eqn:td}. We shift the GLaMM results to dimensional numbers using the numbers from table \ref{tab::LMth} and $\tau_d$. In figure \ref{fig::5} we plot a comparison of the first and third harmonics as a counterpart to figure 4 in the main text. 

The two key observations are (i) that unlike the higher harmonics there is little difference for $De < 100$ for sufficiently entangled melts, $Z \ge 20$ and (ii) the cross-over between $G_3'$ and $G_3''$ occurs after the terminal time cross-over, which is counter to the experimental observations.

\begin{figure}[ht]
	\includegraphics[width=0.47\textwidth ]{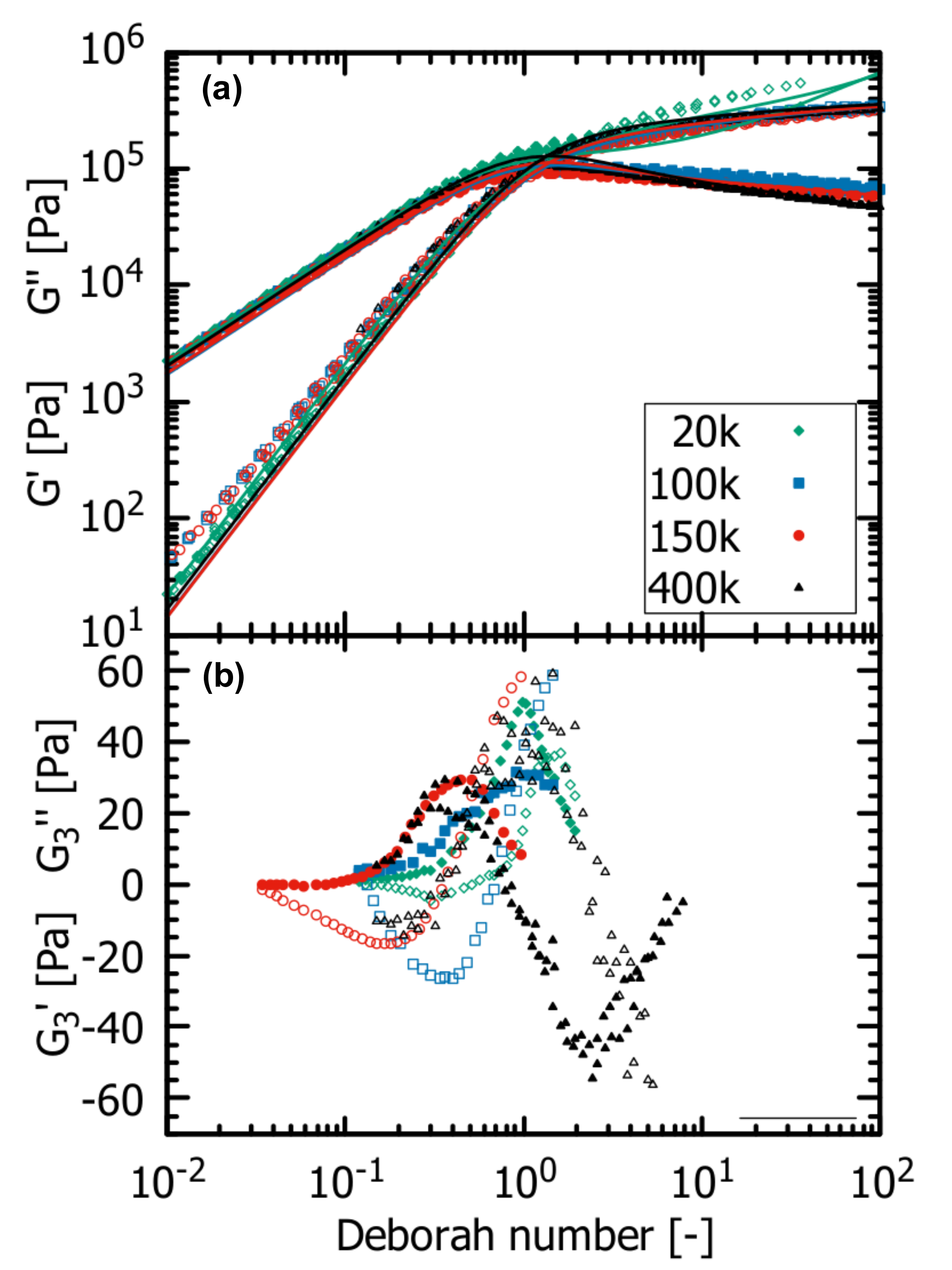}
	\caption{The analogue to figure 1 plotted as a function of Deborah number (frequency normalised by terminal relaxation time $De = \omega\tau_d$). The linear rheology superimposes in the terminal region for both theory and experiment.  \label{fig::6}}
\end{figure}

\begin{figure}[ht]
	\includegraphics[width=0.47\textwidth ]{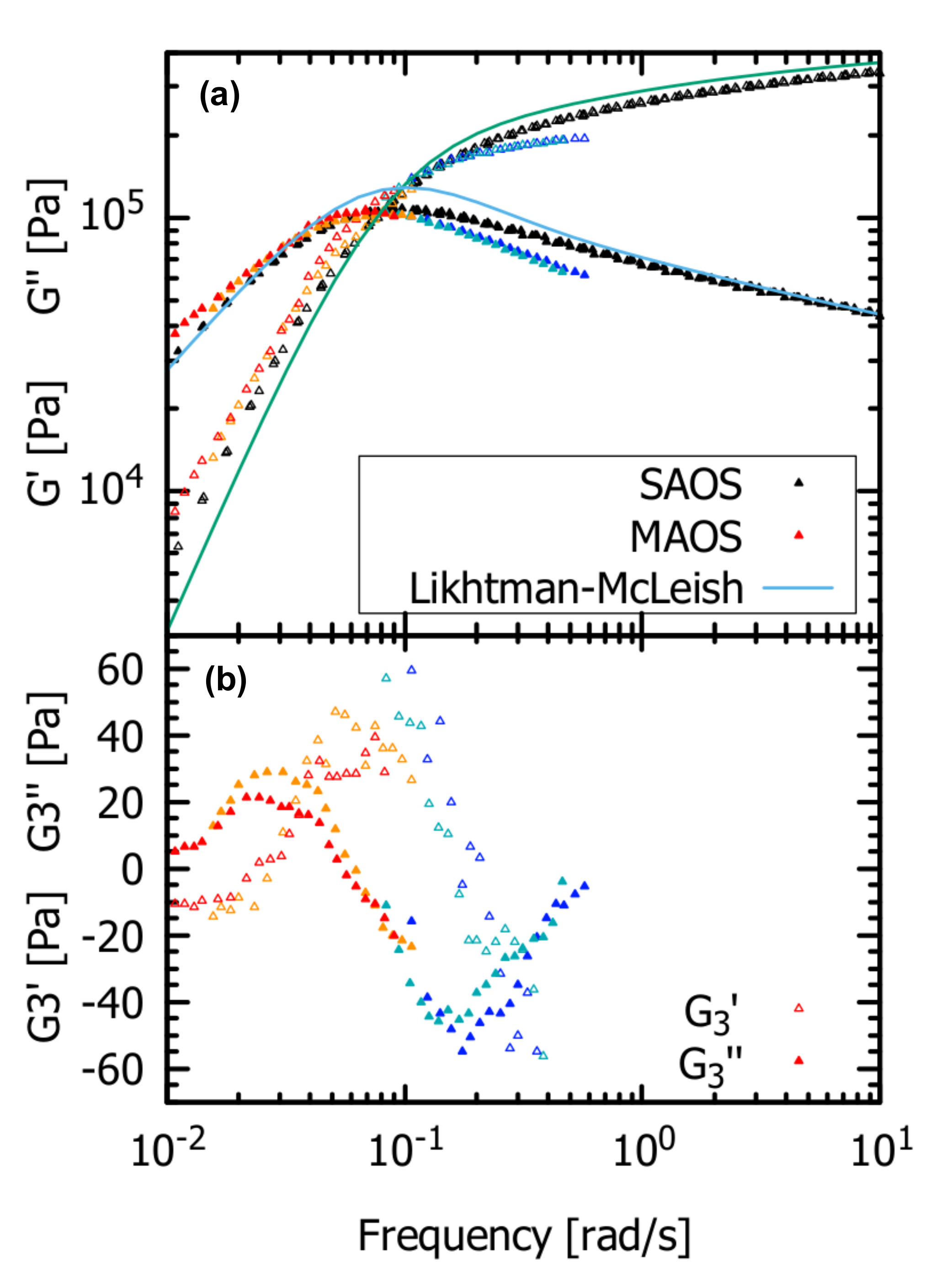}
	\caption{The linear rheology (a) and LAOS results (b) of PI400k at various temperatures. The predictions of the Likhtman-McLeish model are included. Red $50^{\circ}$, Orange $45^{\circ}$, Aqua $30^{\circ}$ and blue $25^{\circ}$. \label{fig::7}}
\end{figure}

\begin{figure}[ht]
	\includegraphics[width=0.99\textwidth]{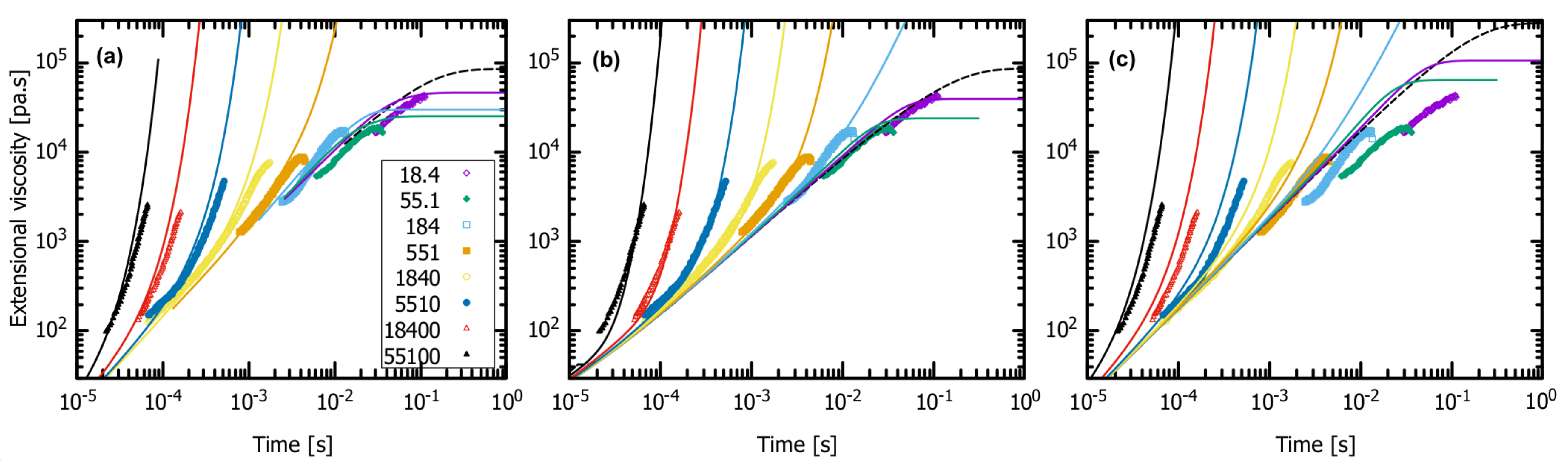}
	\caption{Extensional rheology measured on a SER attachment at $-30^{\circ}$. Each sub-figure compares the extensional data to one of the theories used in the main paper. (a): GLaMM, (b): 9 mode Rolie-Poly, (c): 1 mode Rolie-Poly. \label{fig::8}}
\end{figure}

\begin{figure}[ht]
	\includegraphics[width=0.99\textwidth ]{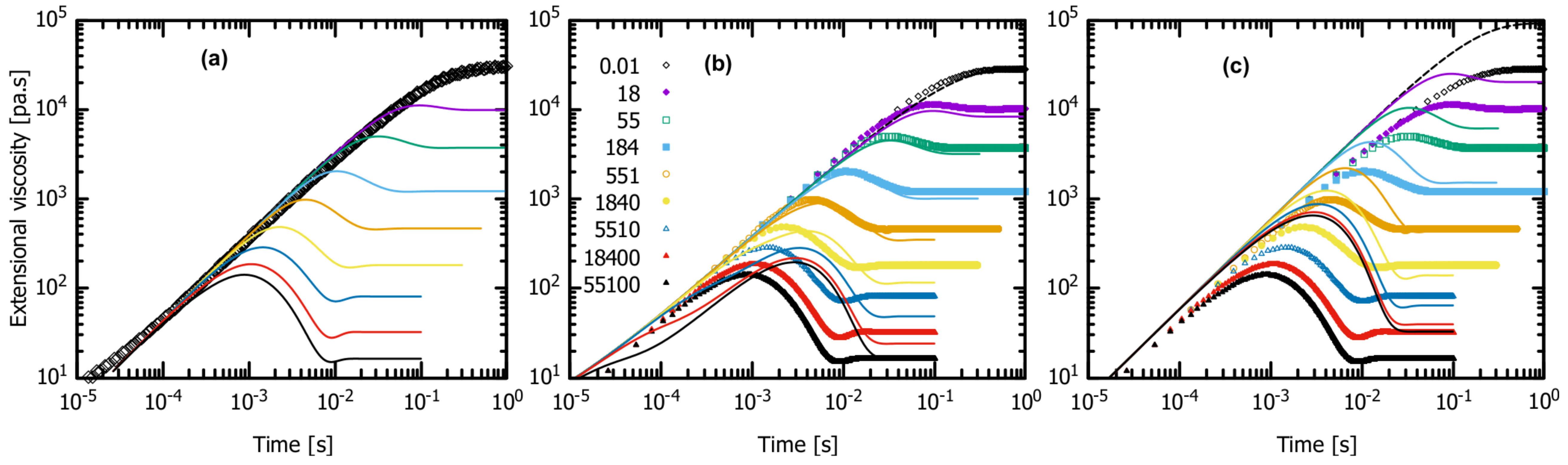}
	\caption{The shear rheology predictions of the three models used in the main paper. (a): the GLaMM model is compared to the linear rheology and the $\eta^*(t)$ envelope can be observed, (b): the 9 mode Rolie-Poly model is compared to GLaMM predictions and (c): the 1 mode Rolie-Poly model is comapred to the GLaMM predictions.  \label{fig::9}}
\end{figure}

Figure \ref{fig::6} shows the equivalent plot for the experimental data, plotted now against Deborah number. For PI400k we used the Williams-Landel-Ferry (WLF) time-temperature superposition (TTS) theory to increase the total flow regime explored. The results for PI400k depicting the various temperatures \ref{fig::7} which were all shifted to $25^{\circ}$C to match the data for other molecular weights.
It should be noted that GLaMM theory for $Z=5$ deviates from the behaviour of PI20k for $De > 1$, which is expected since it is a theory for well entangled melts.

In figure \ref{fig::8} we present a comparison between non-linear extensional experiments and the predictions of the non-linear constitutive theories used in the main paper. We present data for PI100k with measurements taken at $-30^{\circ}C$ and TTS shifted to $25^{\circ}C$. We compare the GLaMM model whose parameters are detailed in tables \ref{tab::LMth} and \ref{tab::GLaMM}, the one mode Rolie-Poly model (table \ref{tab::1RP}) and the 9 mode Rolie-Poly model (table \ref{tab::9RP}).
The GLaMM model (with no free parameters) and the 9-mode RoliePoly model both fit the data excellently. The 1-mode Rolie-Poly model only captures the strain-hardening at the higher rates. In figure \ref{fig::9} we compare the transient shear response of the models. The GLaMM model mimics the linear viscoelastic envelope seen in the extensional data, and in \citet{Auhl2008a-SI} the model is shown to capture non-linear shear response excellently. The 9-mode Rolie-Poly model compares reasonably to the GLaMM model for intermediate rates but deviates from the transient response at higher rates. The 1-mode model fails to capture nearly every aspect of the GLaMM model other than the general form. 

\begin{figure}[ht]
	\includegraphics[width=0.9\textwidth ]{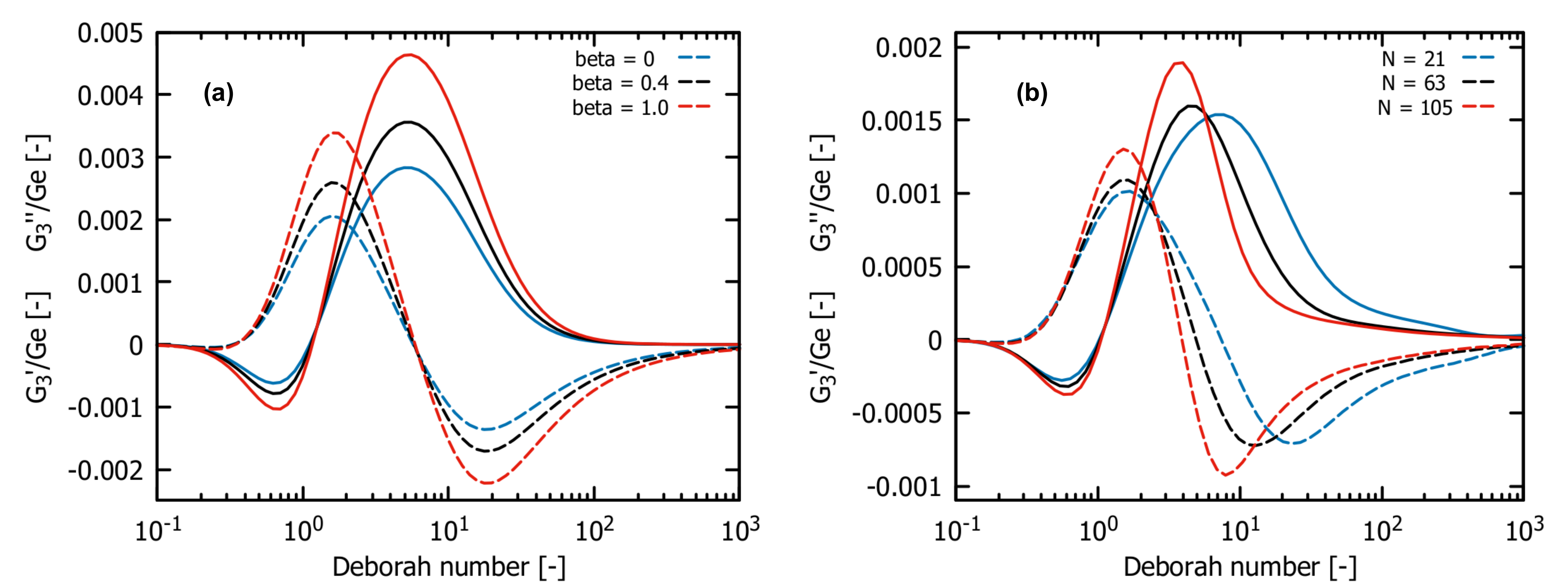}
	\caption{(a): Variations in the third harmonics as a function of the Rolie-Poly parameter $\beta\in[0,1]$. The figure shows that the amplitude of features in the third harmonic increases with increasing $\beta$ which parametrises constraint release. Notice that the cross-over doesn't vary with the amount of CCR. (b): The GLaMM third harmonic predictions shown with different discretisation $N = Z, 3Z, 5Z$. The figure shows little difference in the GLaMM predictions with increased numerical accuracy. \label{fig::10}}
\end{figure}

We then show how the third harmonics vary with some model parameters. In the left figure of \ref{fig::10} we show the Rolie-Poly model (1-mode) for various values of $\beta$, the convective constraint release parameter. The magnitude of the third harmonics increases with increasing CCR, however the frequency dependence is unaffected. The right hand sub-figure of \ref{fig::10} compares the GLaMM model for various values of the discretisation parameter $N$, which is an odd multiple of the entanglement number. Overall, there is little difference for different discretisation numbers, although the magnitude increase with increasing $N$ and the features move to slightly lower $De$. The lowest value $N=Z$ matches experiment with the most accuracy. Also, for $N=105$ simulations took around 3 days per individual frequency to complete and is the practical limit of the simulation time available. For both the above reasons we choose to use $N=Z$ in the main paper. 

\begin{figure}[ht]
	\includegraphics[width=0.6\textwidth ]{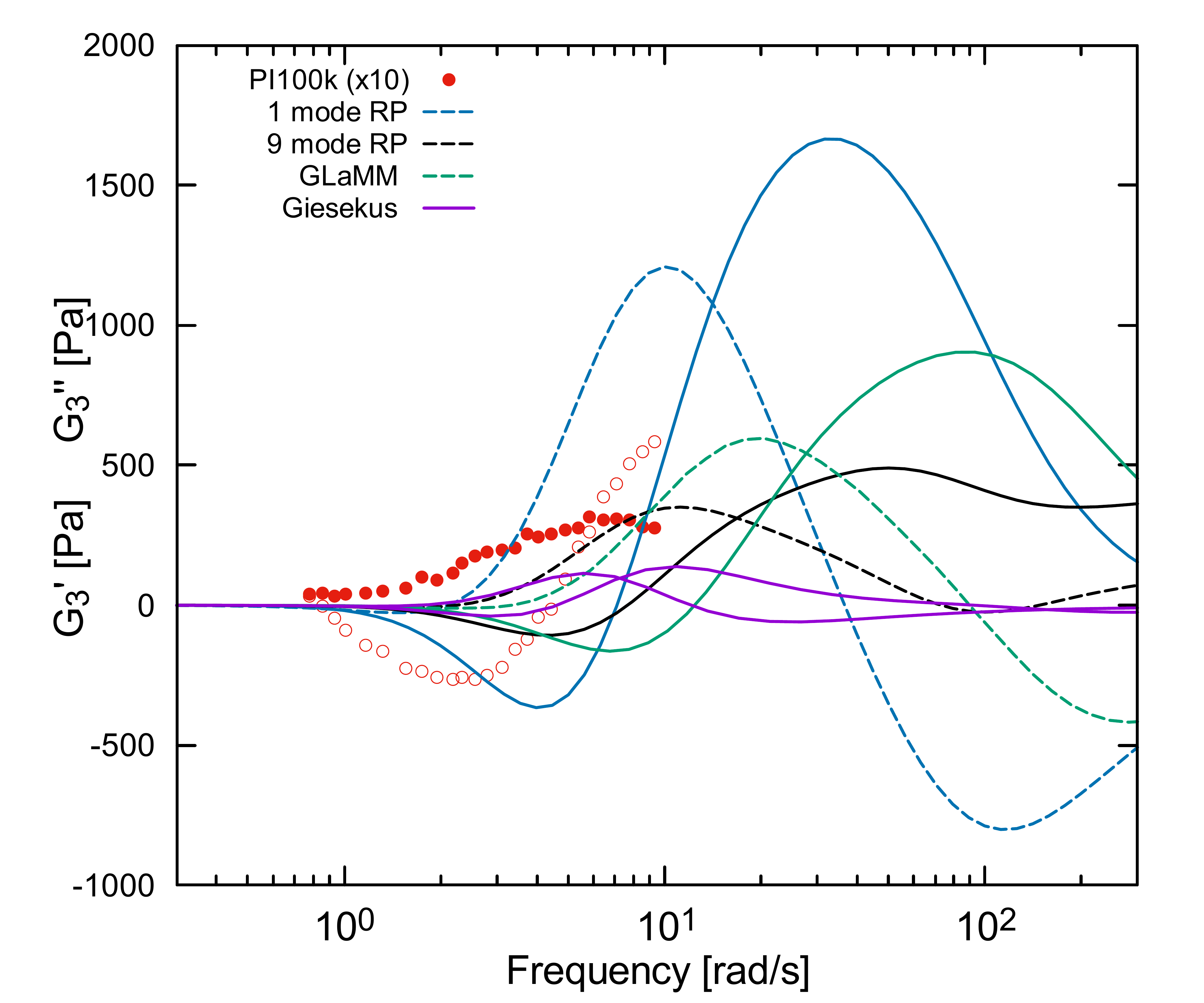}
	\caption{Giesekus predictions for PI100k compared to GlaMM, Rolie-poly and experimental data. The solid lines are $G'_3$ and the dashed lines are for $G''_3$'. \label{fig::11}}
\end{figure}

\begin{figure}[ht]
	\includegraphics[width=0.99\textwidth ]{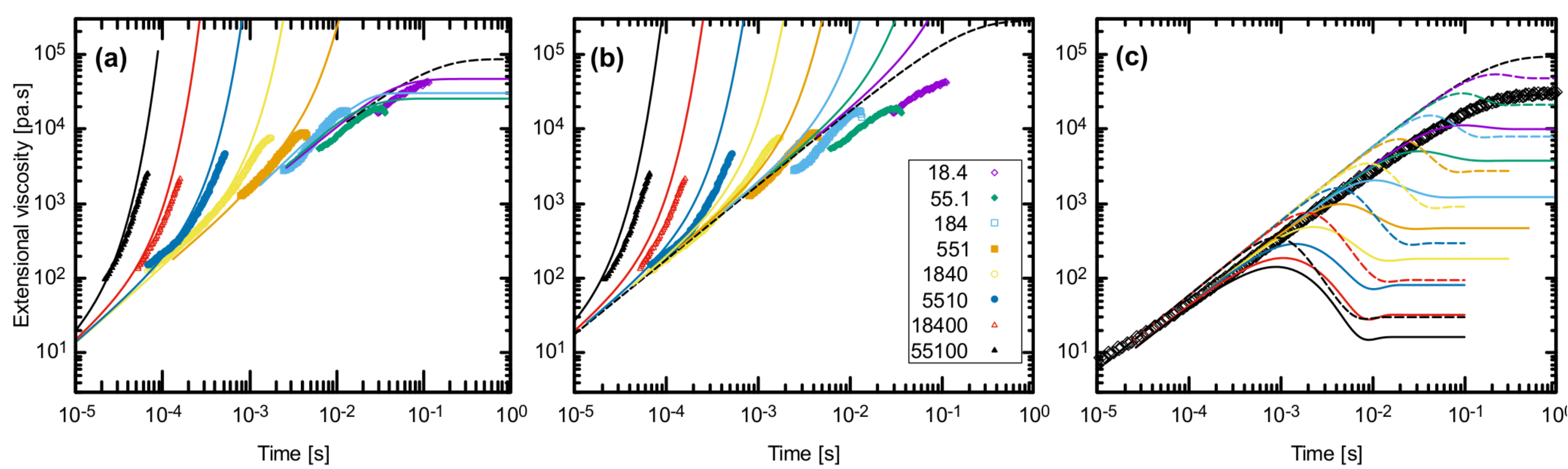}
	\caption{Comparison of experimental data to the shear and extensional rheology predictions of GLaMM and the Giesekus model with alpha = 0.5, showing that the Giesekus model cannot fit all flows simultaneously with a single value of alpha  (a): comparison of GLaMM to experimental extension (b): comparison of Giesekus to experimental extension and (c): GLaMM (solid lines) and Giesekus (dotted lines) compared to experimental transient shear. \label{fig::12}}
\end{figure}

Finally, we include a comparison to a single mode Giesekus \cite{Gurnon2012-SI,Calin2010-SI} prediction. Using a value of $\alpha = 0.5$, this gives slightly lower values of $G_3'$ and $G_3''$ than Rolie-poly and GLaMM (\ref{fig::11}). However, it predicts similar magnitudes for the peaks in $G_3'$ and $G_3''$, whereas the other models and experimental data show $G_3'$ having a significantly larger peak.  Also, crucially, the predictions of transient shear and extensional flows (\ref{fig::12}) are compromised, and it is clear that all 3 flows cannot be captured by the single non-linearity parameter in this model.

\begin{table}[ht] 
	\caption{Rolie-Poly 1 Maxwell mode parameters for PI100k (Z=21) with $\beta=0$.  \label{tab::1RP}} 	\begin{tabular}{ @{\extracolsep{\fill}} c c c }
		\hline 
		\phantom{aaaaa}$G_i$\phantom{aaaaa} & \phantom{aaaa}$\tau_{d_i}$\phantom{aaaa} & \phantom{aaaa}$\tau_{s_i}$\phantom{aaaa}  \\ 
		\hline 
		\hline
		588844  & 1.570E-1 & 0.0057  \\ 
		\hline
	\end{tabular} 
\end{table}

\begin{table}[ht] 
	\caption{Rolie-Poly 9 Maxwell mode parameters for PI100k (Z=21) with $\beta=0$.  \label{tab::9RP}}
	\begin{tabular}{ @{\extracolsep{\fill}} c c c }
		\hline 
		\phantom{aaaaa}$G_i$\phantom{aaaaa} & \phantom{aaaa}$\tau_{d_i}$\phantom{aaaa} & \phantom{aaaa}$\tau_{s_i}$\phantom{aaaa}  \\ 
		\hline 
		\hline
		170661  & 1.493E-1 & 0.0056288  \\ 
		\hline 
		75449.2 & 3.316E-2 & -  \\ 
		\hline 
		69829.9 & 7.365E-3 & -  \\ 
		\hline 
		52453.8 & 1.636E-4 & -  \\ 
		\hline 
		58903.4 & 3.632E-4 & -  \\ 
		\hline 
		66145.8 & 8.066E-5 & -  \\ 
		\hline 
		154820  & 1.791E-5 & -  \\ 
		\hline 
		316517  & 3.978E-6 & -  \\ 
		\hline 
		2103720 & 8.834E-7 & - \\
		\hline
	\end{tabular} 
\end{table}

%


\begin{thebibliography}{44}%
		\makeatletter
		\providecommand \@ifxundefined [1]{%
			\@ifx{#1\undefined}
		}%
		\providecommand \@ifnum [1]{%
			\ifnum #1\expandafter \@firstoftwo
			\else \expandafter \@secondoftwo
			\fi
		}%
		\providecommand \@ifx [1]{%
			\ifx #1\expandafter \@firstoftwo
			\else \expandafter \@secondoftwo
			\fi
		}%
		\providecommand \natexlab [1]{#1}%
		\providecommand \enquote  [1]{``#1''}%
		\providecommand \bibnamefont  [1]{#1}%
		\providecommand \bibfnamefont [1]{#1}%
		\providecommand \citenamefont [1]{#1}%
		\providecommand \href@noop [0]{\@secondoftwo}%
		\providecommand \href [0]{\begingroup \@sanitize@url \@href}%
		\providecommand \@href[1]{\@@startlink{#1}\@@href}%
		\providecommand \@@href[1]{\endgroup#1\@@endlink}%
		\providecommand \@sanitize@url [0]{\catcode `\\12\catcode `\$12\catcode
			`\&12\catcode `\#12\catcode `\^12\catcode `\_12\catcode `\%12\relax}%
		\providecommand \@@startlink[1]{}%
		\providecommand \@@endlink[0]{}%
		\providecommand \url  [0]{\begingroup\@sanitize@url \@url }%
		\providecommand \@url [1]{\endgroup\@href {#1}{\urlprefix }}%
		\providecommand \urlprefix  [0]{URL }%
		\providecommand \Eprint [0]{\href }%
		\providecommand \doibase [0]{http://dx.doi.org/}%
		\providecommand \selectlanguage [0]{\@gobble}%
		\providecommand \bibinfo  [0]{\@secondoftwo}%
		\providecommand \bibfield  [0]{\@secondoftwo}%
		\providecommand \translation [1]{[#1]}%
		\providecommand \BibitemOpen [0]{}%
		\providecommand \bibitemStop [0]{}%
		\providecommand \bibitemNoStop [0]{.\EOS\space}%
		\providecommand \EOS [0]{\spacefactor3000\relax}%
		\providecommand \BibitemShut  [1]{\csname bibitem#1\endcsname}%
		\let\auto@bib@innerbib\@empty
		\bibitem [{\citenamefont {Malmberg}\ \emph {et~al.}(2002)\citenamefont
			{Malmberg}, \citenamefont {Gabriel}, \citenamefont {Steffl}, \citenamefont
			{M{\"{u}}nstedt},\ and\ \citenamefont {L{\"{o}}fgren}}]{Malmberg2002}%
		\BibitemOpen
		\bibfield  {author} {\bibinfo {author} {\bibfnamefont {A.}~\bibnamefont
				{Malmberg}}, \bibinfo {author} {\bibfnamefont {C.}~\bibnamefont {Gabriel}},
			\bibinfo {author} {\bibfnamefont {T.}~\bibnamefont {Steffl}}, \bibinfo
			{author} {\bibfnamefont {H.}~\bibnamefont {M{\"{u}}nstedt}}, \ and\ \bibinfo
			{author} {\bibfnamefont {B.}~\bibnamefont {L{\"{o}}fgren}},\ }\href {\doibase
			10.1021/ma010753l} {\bibfield  {journal} {\bibinfo  {journal}
				{Macromolecules}\ }\textbf {\bibinfo {volume} {35}},\ \bibinfo {pages} {1038}
			(\bibinfo {year} {2002})}\BibitemShut {NoStop}%
		\bibitem [{\citenamefont {Vega}\ \emph {et~al.}(1998)\citenamefont {Vega},
			\citenamefont {Santamaria}, \citenamefont {Munoz-Escalona},\ and\
			\citenamefont {Lafuente}}]{Vega1998}%
		\BibitemOpen
		\bibfield  {author} {\bibinfo {author} {\bibfnamefont {J.~F.}\ \bibnamefont
				{Vega}}, \bibinfo {author} {\bibfnamefont {A.}~\bibnamefont {Santamaria}},
			\bibinfo {author} {\bibfnamefont {A.}~\bibnamefont {Munoz-Escalona}}, \ and\
			\bibinfo {author} {\bibfnamefont {P.}~\bibnamefont {Lafuente}},\ }\href
		{\doibase 10.1021/ma9708961} {\bibfield  {journal} {\bibinfo  {journal}
				{Macromolecules}\ }\textbf {\bibinfo {volume} {31}},\ \bibinfo {pages} {3639}
			(\bibinfo {year} {1998})}\BibitemShut {NoStop}%
		\bibitem [{\citenamefont {Vega}\ \emph {et~al.}(1999)\citenamefont {Vega},
			\citenamefont {Fernandez},\ and\ \citenamefont {Santamaria}}]{Vega1999}%
		\BibitemOpen
		\bibfield  {author} {\bibinfo {author} {\bibfnamefont {J.~F.}\ \bibnamefont
				{Vega}}, \bibinfo {author} {\bibfnamefont {M.}~\bibnamefont {Fernandez}}, \
			and\ \bibinfo {author} {\bibfnamefont {A.}~\bibnamefont {Santamaria}},\
		}\href {\doibase
			10.1002/(SICI)1521-3935(19991001)200:10<2257::AID-MACP2257>3.0.CO;2-L}
		{\bibfield  {journal} {\bibinfo  {journal} {Macromolecular}\ }\textbf
			{\bibinfo {volume} {2268}},\ \bibinfo {pages} {2257} (\bibinfo {year}
			{1999})}\BibitemShut {NoStop}%
		\bibitem [{\citenamefont {Zhang}\ \emph {et~al.}(2008)\citenamefont {Zhang},
			\citenamefont {Cheng}, \citenamefont {Zhang}, \citenamefont {Sun},
			\citenamefont {Dong},\ and\ \citenamefont {Han}}]{Zhang2008a}%
		\BibitemOpen
		\bibfield  {author} {\bibinfo {author} {\bibfnamefont {R.}~\bibnamefont
				{Zhang}}, \bibinfo {author} {\bibfnamefont {H.}~\bibnamefont {Cheng}},
			\bibinfo {author} {\bibfnamefont {C.}~\bibnamefont {Zhang}}, \bibinfo
			{author} {\bibfnamefont {T.}~\bibnamefont {Sun}}, \bibinfo {author}
			{\bibfnamefont {X.}~\bibnamefont {Dong}}, \ and\ \bibinfo {author}
			{\bibfnamefont {C.~C.}\ \bibnamefont {Han}},\ }\href {\doibase
			10.1021/ma800646s} {\bibfield  {journal} {\bibinfo  {journal}
				{Macromolecules}\ }\textbf {\bibinfo {volume} {41}},\ \bibinfo {pages} {6818}
			(\bibinfo {year} {2008})}\BibitemShut {NoStop}%
		\bibitem [{\citenamefont {Hyun}\ \emph {et~al.}(2011)\citenamefont {Hyun},
			\citenamefont {Wilhelm}, \citenamefont {Klein}, \citenamefont {Cho},
			\citenamefont {Nam}, \citenamefont {Ahn}, \citenamefont {Lee}, \citenamefont
			{Ewoldt},\ and\ \citenamefont {McKinley}}]{Hyun2011}%
		\BibitemOpen
		\bibfield  {author} {\bibinfo {author} {\bibfnamefont {K.}~\bibnamefont
				{Hyun}}, \bibinfo {author} {\bibfnamefont {M.}~\bibnamefont {Wilhelm}},
			\bibinfo {author} {\bibfnamefont {C.~O.}\ \bibnamefont {Klein}}, \bibinfo
			{author} {\bibfnamefont {K.~S.}\ \bibnamefont {Cho}}, \bibinfo {author}
			{\bibfnamefont {J.~G.}\ \bibnamefont {Nam}}, \bibinfo {author} {\bibfnamefont
				{K.~H.}\ \bibnamefont {Ahn}}, \bibinfo {author} {\bibfnamefont {S.~J.}\
				\bibnamefont {Lee}}, \bibinfo {author} {\bibfnamefont {R.~H.}\ \bibnamefont
				{Ewoldt}}, \ and\ \bibinfo {author} {\bibfnamefont {G.~H.}\ \bibnamefont
				{McKinley}},\ }\href {\doibase 10.1016/j.progpolymsci.2011.02.002} {\bibfield
			{journal} {\bibinfo  {journal} {Progress in Polymer Science}\ }\textbf
			{\bibinfo {volume} {36}},\ \bibinfo {pages} {1697} (\bibinfo {year}
			{2011})}\BibitemShut {NoStop}%
		\bibitem [{\citenamefont {Ng}\ \emph {et~al.}(2011)\citenamefont {Ng},
			\citenamefont {McKinley},\ and\ \citenamefont {Ewoldt}}]{Ng2011}%
		\BibitemOpen
		\bibfield  {author} {\bibinfo {author} {\bibfnamefont {T.~S.~K.}\
				\bibnamefont {Ng}}, \bibinfo {author} {\bibfnamefont {G.~H.}\ \bibnamefont
				{McKinley}}, \ and\ \bibinfo {author} {\bibfnamefont {R.~H.}\ \bibnamefont
				{Ewoldt}},\ }\href {\doibase 10.1122/1.3570340} {\bibfield  {journal}
			{\bibinfo  {journal} {Journal of Rheology}\ }\textbf {\bibinfo {volume}
				{55}},\ \bibinfo {pages} {627} (\bibinfo {year} {2011})}\BibitemShut
		{NoStop}%
		\bibitem [{\citenamefont {Martinetti}\ \emph {et~al.}(2014)\citenamefont
			{Martinetti}, \citenamefont {Mannion}, \citenamefont {Voje}, \citenamefont
			{Xie}, \citenamefont {Ewoldt}, \citenamefont {Morgret}, \citenamefont
			{Bates},\ and\ \citenamefont {Macosko}}]{Martinetti2014}%
		\BibitemOpen
		\bibfield  {author} {\bibinfo {author} {\bibfnamefont {L.}~\bibnamefont
				{Martinetti}}, \bibinfo {author} {\bibfnamefont {A.~M.}\ \bibnamefont
				{Mannion}}, \bibinfo {author} {\bibfnamefont {W.~E.}\ \bibnamefont {Voje}},
			\bibinfo {author} {\bibfnamefont {R.}~\bibnamefont {Xie}}, \bibinfo {author}
			{\bibfnamefont {R.~H.}\ \bibnamefont {Ewoldt}}, \bibinfo {author}
			{\bibfnamefont {L.~D.}\ \bibnamefont {Morgret}}, \bibinfo {author}
			{\bibfnamefont {F.~S.}\ \bibnamefont {Bates}}, \ and\ \bibinfo {author}
			{\bibfnamefont {C.~W.}\ \bibnamefont {Macosko}},\ }\href {\doibase
			10.1122/1.4874322} {\bibfield  {journal} {\bibinfo  {journal} {Journal of
					Rheology}\ }\textbf {\bibinfo {volume} {58}},\ \bibinfo {pages} {821}
			(\bibinfo {year} {2014})}\BibitemShut {NoStop}%
		\bibitem [{\citenamefont {Bharadwaj}\ \emph {et~al.}(2017)\citenamefont
			{Bharadwaj}, \citenamefont {Schweizer},\ and\ \citenamefont
			{Ewoldt}}]{Bharadwaj2017}%
		\BibitemOpen
		\bibfield  {author} {\bibinfo {author} {\bibfnamefont {N.}~\bibnamefont
				{Bharadwaj}}, \bibinfo {author} {\bibfnamefont {K.}~\bibnamefont
				{Schweizer}}, \ and\ \bibinfo {author} {\bibfnamefont {R.~H.}\ \bibnamefont
				{Ewoldt}},\ }\href {\doibase 10.1122/1.4979368} {\bibfield  {journal}
			{\bibinfo  {journal} {Journal of Rheology}\ }\textbf {\bibinfo {volume}
				{643}} (\bibinfo {year} {2017}),\ 10.1122/1.4979368}\BibitemShut {NoStop}%
		\bibitem [{\citenamefont {Dimitriou}\ \emph {et~al.}(2012)\citenamefont
			{Dimitriou}, \citenamefont {Casanellas}, \citenamefont {Ober},\ and\
			\citenamefont {McKinley}}]{Dimitriou2012}%
		\BibitemOpen
		\bibfield  {author} {\bibinfo {author} {\bibfnamefont {C.~J.}\ \bibnamefont
				{Dimitriou}}, \bibinfo {author} {\bibfnamefont {L.}~\bibnamefont
				{Casanellas}}, \bibinfo {author} {\bibfnamefont {T.~J.}\ \bibnamefont
				{Ober}}, \ and\ \bibinfo {author} {\bibfnamefont {G.~H.}\ \bibnamefont
				{McKinley}},\ }\href {\doibase 10.1007/s00397-012-0619-9} {\bibfield
			{journal} {\bibinfo  {journal} {Rheologica Acta}\ }\textbf {\bibinfo {volume}
				{51}},\ \bibinfo {pages} {395} (\bibinfo {year} {2012})}\BibitemShut
		{NoStop}%
		\bibitem [{\citenamefont {Renou}\ \emph {et~al.}(2010)\citenamefont {Renou},
			\citenamefont {Stellbrink},\ and\ \citenamefont {Petekidis}}]{Renou2010}%
		\BibitemOpen
		\bibfield  {author} {\bibinfo {author} {\bibfnamefont {F.}~\bibnamefont
				{Renou}}, \bibinfo {author} {\bibfnamefont {J.}~\bibnamefont {Stellbrink}}, \
			and\ \bibinfo {author} {\bibfnamefont {G.}~\bibnamefont {Petekidis}},\ }\href
		{\doibase 10.1122/1.3483610} {\bibfield  {journal} {\bibinfo  {journal}
				{Journal of Rheology}\ }\textbf {\bibinfo {volume} {54}},\ \bibinfo {pages}
			{1219} (\bibinfo {year} {2010})}\BibitemShut {NoStop}%
		\bibitem [{\citenamefont {Duvarci}\ \emph {et~al.}(2017)\citenamefont
			{Duvarci}, \citenamefont {Yazar},\ and\ \citenamefont
			{Kokini}}]{Duvarci2017}%
		\BibitemOpen
		\bibfield  {author} {\bibinfo {author} {\bibfnamefont {O.~C.}\ \bibnamefont
				{Duvarci}}, \bibinfo {author} {\bibfnamefont {G.}~\bibnamefont {Yazar}}, \
			and\ \bibinfo {author} {\bibfnamefont {J.~L.}\ \bibnamefont {Kokini}},\
		}\href {\doibase 10.1016/j.tifs.2016.08.014} {\bibfield  {journal} {\bibinfo
				{journal} {Trends in Food Science {\&} Technology}\ }\textbf {\bibinfo
				{volume} {60}},\ \bibinfo {pages} {2} (\bibinfo {year} {2017})}\BibitemShut
		{NoStop}%
		\bibitem [{\citenamefont {Li}\ \emph {et~al.}(2009)\citenamefont {Li},
			\citenamefont {Wang},\ and\ \citenamefont {Wang}}]{Li2009}%
		\BibitemOpen
		\bibfield  {author} {\bibinfo {author} {\bibfnamefont {X.}~\bibnamefont
				{Li}}, \bibinfo {author} {\bibfnamefont {S.-Q.}\ \bibnamefont {Wang}}, \ and\
			\bibinfo {author} {\bibfnamefont {X.}~\bibnamefont {Wang}},\ }\href {\doibase
			10.1122/1.3193713} {\enquote {\bibinfo {title} {{Nonlinearity in large
						amplitude oscillatory shear (LAOS) of different viscoelastic materials}},}\ }
		(\bibinfo {year} {2009})\BibitemShut {NoStop}%
		\bibitem [{\citenamefont {Ewoldt}\ \emph {et~al.}(2008)\citenamefont {Ewoldt},
			\citenamefont {Hosoi},\ and\ \citenamefont {McKinley}}]{Ewoldt2008}%
		\BibitemOpen
		\bibfield  {author} {\bibinfo {author} {\bibfnamefont {R.~H.}\ \bibnamefont
				{Ewoldt}}, \bibinfo {author} {\bibfnamefont {A.~E.}\ \bibnamefont {Hosoi}}, \
			and\ \bibinfo {author} {\bibfnamefont {G.~H.}\ \bibnamefont {McKinley}},\
		}\href {\doibase 10.1122/1.2970095} {\bibfield  {journal} {\bibinfo
				{journal} {Journal of Rheology}\ }\textbf {\bibinfo {volume} {52}},\ \bibinfo
			{pages} {1427} (\bibinfo {year} {2008})}\BibitemShut {NoStop}%
		\bibitem [{\citenamefont {Cziep}\ \emph {et~al.}(2016)\citenamefont {Cziep},
			\citenamefont {Abbasi}, \citenamefont {Heck}, \citenamefont {Arens},\ and\
			\citenamefont {Wilhelm}}]{Cziep2016}%
		\BibitemOpen
		\bibfield  {author} {\bibinfo {author} {\bibfnamefont {M.~A.}\ \bibnamefont
				{Cziep}}, \bibinfo {author} {\bibfnamefont {M.}~\bibnamefont {Abbasi}},
			\bibinfo {author} {\bibfnamefont {M.}~\bibnamefont {Heck}}, \bibinfo {author}
			{\bibfnamefont {L.}~\bibnamefont {Arens}}, \ and\ \bibinfo {author}
			{\bibfnamefont {M.}~\bibnamefont {Wilhelm}},\ }\href {\doibase
			10.1021/acs.macromol.5b02706} {\bibfield  {journal} {\bibinfo  {journal}
				{Macromolecules}\ }\textbf {\bibinfo {volume} {49}},\ \bibinfo {pages} {3566}
			(\bibinfo {year} {2016})}\BibitemShut {NoStop}%
		\bibitem [{\citenamefont {Song}\ \emph {et~al.}(2016)\citenamefont {Song},
			\citenamefont {Nnyigide}, \citenamefont {Salehiyan},\ and\ \citenamefont
			{Hyun}}]{Song2016}%
		\BibitemOpen
		\bibfield  {author} {\bibinfo {author} {\bibfnamefont {H.~Y.}\ \bibnamefont
				{Song}}, \bibinfo {author} {\bibfnamefont {O.~S.}\ \bibnamefont {Nnyigide}},
			\bibinfo {author} {\bibfnamefont {R.}~\bibnamefont {Salehiyan}}, \ and\
			\bibinfo {author} {\bibfnamefont {K.}~\bibnamefont {Hyun}},\ }\href {\doibase
			10.1016/j.polymer.2016.04.052} {\bibfield  {journal} {\bibinfo  {journal}
				{Polymer}\ }\textbf {\bibinfo {volume} {104}},\ \bibinfo {pages} {268}
			(\bibinfo {year} {2016})}\BibitemShut {NoStop}%
		\bibitem [{\citenamefont {Wagner}\ \emph {et~al.}(2011)\citenamefont {Wagner},
			\citenamefont {Rol{\'{o}}n-Garrido}, \citenamefont {Hyun},\ and\
			\citenamefont {Wilhelm}}]{Wagner2011}%
		\BibitemOpen
		\bibfield  {author} {\bibinfo {author} {\bibfnamefont {M.~H.}\ \bibnamefont
				{Wagner}}, \bibinfo {author} {\bibfnamefont {V.~H.}\ \bibnamefont
				{Rol{\'{o}}n-Garrido}}, \bibinfo {author} {\bibfnamefont {K.}~\bibnamefont
				{Hyun}}, \ and\ \bibinfo {author} {\bibfnamefont {M.}~\bibnamefont
				{Wilhelm}},\ }\href {\doibase 10.1122/1.3553031} {\bibfield  {journal}
			{\bibinfo  {journal} {Journal of Rheology}\ }\textbf {\bibinfo {volume}
				{55}},\ \bibinfo {pages} {495} (\bibinfo {year} {2011})}\BibitemShut
		{NoStop}%
		\bibitem [{\citenamefont {Hoyle}\ \emph {et~al.}(2014)\citenamefont {Hoyle},
			\citenamefont {Auhl}, \citenamefont {Harlen}, \citenamefont {Barroso},
			\citenamefont {Wilhelm},\ and\ \citenamefont {McLeish}}]{Hoyle2014}%
		\BibitemOpen
		\bibfield  {author} {\bibinfo {author} {\bibfnamefont {D.~M.}\ \bibnamefont
				{Hoyle}}, \bibinfo {author} {\bibfnamefont {D.~W.}\ \bibnamefont {Auhl}},
			\bibinfo {author} {\bibfnamefont {O.~G.}\ \bibnamefont {Harlen}}, \bibinfo
			{author} {\bibfnamefont {V.~C.}\ \bibnamefont {Barroso}}, \bibinfo {author}
			{\bibfnamefont {M.}~\bibnamefont {Wilhelm}}, \ and\ \bibinfo {author}
			{\bibfnamefont {T.~C.~B.}\ \bibnamefont {McLeish}},\ }\href {\doibase
			10.1122/1.4881467} {\bibfield  {journal} {\bibinfo  {journal} {Journal of
					Rheology}\ }\textbf {\bibinfo {volume} {58}},\ \bibinfo {pages} {969}
			(\bibinfo {year} {2014})}\BibitemShut {NoStop}%
		\bibitem [{\citenamefont {Vittorias}\ \emph {et~al.}(2006)\citenamefont
			{Vittorias}, \citenamefont {Parkinson}, \citenamefont {Klimke}, \citenamefont
			{Debbaut},\ and\ \citenamefont {Wilhelm}}]{Vittorias2006}%
		\BibitemOpen
		\bibfield  {author} {\bibinfo {author} {\bibfnamefont {I.}~\bibnamefont
				{Vittorias}}, \bibinfo {author} {\bibfnamefont {M.}~\bibnamefont
				{Parkinson}}, \bibinfo {author} {\bibfnamefont {K.}~\bibnamefont {Klimke}},
			\bibinfo {author} {\bibfnamefont {B.}~\bibnamefont {Debbaut}}, \ and\
			\bibinfo {author} {\bibfnamefont {M.}~\bibnamefont {Wilhelm}},\ }\href
		{\doibase 10.1007/s00397-006-0111-5} {\bibfield  {journal} {\bibinfo
				{journal} {Rheologica Acta}\ }\textbf {\bibinfo {volume} {46}},\ \bibinfo
			{pages} {321} (\bibinfo {year} {2006})}\BibitemShut {NoStop}%
		\bibitem [{\citenamefont {Abbasi}\ \emph {et~al.}(2013)\citenamefont {Abbasi},
			\citenamefont {{Golshan Ebrahimi}},\ and\ \citenamefont
			{Wilhelm}}]{Abbasi2013}%
		\BibitemOpen
		\bibfield  {author} {\bibinfo {author} {\bibfnamefont {M.}~\bibnamefont
				{Abbasi}}, \bibinfo {author} {\bibfnamefont {N.}~\bibnamefont {{Golshan
						Ebrahimi}}}, \ and\ \bibinfo {author} {\bibfnamefont {M.}~\bibnamefont
				{Wilhelm}},\ }\href {\doibase 10.1122/1.4824364} {\bibfield  {journal}
			{\bibinfo  {journal} {Journal of Rheology}\ }\textbf {\bibinfo {volume}
				{57}},\ \bibinfo {pages} {1693} (\bibinfo {year} {2013})}\BibitemShut
		{NoStop}%
		\bibitem [{\citenamefont {Clemeur}\ \emph {et~al.}(2003)\citenamefont
			{Clemeur}, \citenamefont {Rutgers},\ and\ \citenamefont
			{Debbaut}}]{Clemeur2003}%
		\BibitemOpen
		\bibfield  {author} {\bibinfo {author} {\bibfnamefont {N.}~\bibnamefont
				{Clemeur}}, \bibinfo {author} {\bibfnamefont {R.~P.~G.}\ \bibnamefont
				{Rutgers}}, \ and\ \bibinfo {author} {\bibfnamefont {B.}~\bibnamefont
				{Debbaut}},\ }\href {\doibase 10.1007/s00397-002-0279-2} {\bibfield
			{journal} {\bibinfo  {journal} {Rheologica acta}\ }\textbf {\bibinfo {volume}
				{42}},\ \bibinfo {pages} {217} (\bibinfo {year} {2003})}\BibitemShut
		{NoStop}%
		\bibitem [{\citenamefont {Giacomin}\ and\ \citenamefont
			{Oakley}(1993)}]{Giacomin1993a}%
		\BibitemOpen
		\bibfield  {author} {\bibinfo {author} {\bibfnamefont {A.}~\bibnamefont
				{Giacomin}}\ and\ \bibinfo {author} {\bibfnamefont {J.}~\bibnamefont
				{Oakley}},\ }\href {http://link.springer.com/article/10.1007/BF00434197}
		{\bibfield  {journal} {\bibinfo  {journal} {Rheologica acta}\ }\textbf
			{\bibinfo {volume} {32}},\ \bibinfo {pages} {328} (\bibinfo {year}
			{1993})}\BibitemShut {NoStop}%
		\bibitem [{\citenamefont {Ewoldt}\ and\ \citenamefont
			{McKinley}(2010)}]{Ewoldt2010}%
		\BibitemOpen
		\bibfield  {author} {\bibinfo {author} {\bibfnamefont {R.~H.}\ \bibnamefont
				{Ewoldt}}\ and\ \bibinfo {author} {\bibfnamefont {G.~H.}\ \bibnamefont
				{McKinley}},\ }\href {\doibase 10.1007/s00397-009-0408-2} {\bibfield
			{journal} {\bibinfo  {journal} {Rheologica Acta}\ }\textbf {\bibinfo {volume}
				{49}},\ \bibinfo {pages} {213} (\bibinfo {year} {2010})}\BibitemShut
		{NoStop}%
		\bibitem [{\citenamefont {Cho}\ \emph {et~al.}(2005)\citenamefont {Cho},
			\citenamefont {Hyun}, \citenamefont {Ahn},\ and\ \citenamefont
			{Lee}}]{Cho2005}%
		\BibitemOpen
		\bibfield  {author} {\bibinfo {author} {\bibfnamefont {K.~S.}\ \bibnamefont
				{Cho}}, \bibinfo {author} {\bibfnamefont {K.}~\bibnamefont {Hyun}}, \bibinfo
			{author} {\bibfnamefont {K.~H.}\ \bibnamefont {Ahn}}, \ and\ \bibinfo
			{author} {\bibfnamefont {S.~J.}\ \bibnamefont {Lee}},\ }\href {\doibase
			10.1122/1.1895801} {\bibfield  {journal} {\bibinfo  {journal} {Journal of
					Rheology}\ }\textbf {\bibinfo {volume} {49}},\ \bibinfo {pages} {747}
			(\bibinfo {year} {2005})}\BibitemShut {NoStop}%
		\bibitem [{\citenamefont {MacSporran}\ and\ \citenamefont
			{Spiers}(1984)}]{MacSporran1984}%
		\BibitemOpen
		\bibfield  {author} {\bibinfo {author} {\bibfnamefont {W.~C.}\ \bibnamefont
				{MacSporran}}\ and\ \bibinfo {author} {\bibfnamefont {R.~P.}\ \bibnamefont
				{Spiers}},\ }\href {\doibase 10.1007/BF01333880} {\bibfield  {journal}
			{\bibinfo  {journal} {Rheologica Acta}\ }\textbf {\bibinfo {volume} {23}},\
			\bibinfo {pages} {90} (\bibinfo {year} {1984})}\BibitemShut {NoStop}%
		\bibitem [{\citenamefont {Wilhelm}\ \emph {et~al.}(1998)\citenamefont
			{Wilhelm}, \citenamefont {Maring},\ and\ \citenamefont
			{Spiess}}]{Wilhelm1998}%
		\BibitemOpen
		\bibfield  {author} {\bibinfo {author} {\bibfnamefont {M.}~\bibnamefont
				{Wilhelm}}, \bibinfo {author} {\bibfnamefont {D.}~\bibnamefont {Maring}}, \
			and\ \bibinfo {author} {\bibfnamefont {H.-W.}\ \bibnamefont {Spiess}},\
		}\href {\doibase 10.1007/s003970050126} {\bibfield  {journal} {\bibinfo
				{journal} {Rheologica Acta}\ }\textbf {\bibinfo {volume} {37}},\ \bibinfo
			{pages} {399} (\bibinfo {year} {1998})}\BibitemShut {NoStop}%
		\bibitem [{\citenamefont {Wilhelm}\ \emph {et~al.}(1999)\citenamefont
			{Wilhelm}, \citenamefont {Reinheimer},\ and\ \citenamefont
			{Ortseifer}}]{Wilhelm1999}%
		\BibitemOpen
		\bibfield  {author} {\bibinfo {author} {\bibfnamefont {M.}~\bibnamefont
				{Wilhelm}}, \bibinfo {author} {\bibfnamefont {P.}~\bibnamefont {Reinheimer}},
			\ and\ \bibinfo {author} {\bibfnamefont {M.}~\bibnamefont {Ortseifer}},\
		}\href {\doibase 10.1007/s003970050185} {\bibfield  {journal} {\bibinfo
				{journal} {Rheologica Acta}\ }\textbf {\bibinfo {volume} {38}},\ \bibinfo
			{pages} {349} (\bibinfo {year} {1999})}\BibitemShut {NoStop}%
		\bibitem [{\citenamefont {Wilhelm}(2002)}]{Wilhelm2002}%
		\BibitemOpen
		\bibfield  {author} {\bibinfo {author} {\bibfnamefont {M.}~\bibnamefont
				{Wilhelm}},\ }\href {\doibase
			10.1002/1439-2054(20020201)287:2<83::AID-MAME83>3.0.CO;2-B} {\bibfield
			{journal} {\bibinfo  {journal} {Macromolecular Materials and Engineering}\
			}\textbf {\bibinfo {volume} {287}},\ \bibinfo {pages} {83} (\bibinfo {year}
			{2002})}\BibitemShut {NoStop}%
		\bibitem [{\citenamefont {Nam}\ \emph {et~al.}(2008)\citenamefont {Nam},
			\citenamefont {Hyun}, \citenamefont {Ahn},\ and\ \citenamefont
			{Lee}}]{Nam2008}%
		\BibitemOpen
		\bibfield  {author} {\bibinfo {author} {\bibfnamefont {J.~G.}\ \bibnamefont
				{Nam}}, \bibinfo {author} {\bibfnamefont {K.}~\bibnamefont {Hyun}}, \bibinfo
			{author} {\bibfnamefont {K.~H.}\ \bibnamefont {Ahn}}, \ and\ \bibinfo
			{author} {\bibfnamefont {S.~J.}\ \bibnamefont {Lee}},\ }\href {\doibase
			10.1016/j.jnnfm.2007.10.002} {\bibfield  {journal} {\bibinfo  {journal}
				{Journal of Non-Newtonian Fluid Mechanics}\ }\textbf {\bibinfo {volume}
				{150}},\ \bibinfo {pages} {1} (\bibinfo {year} {2008})}\BibitemShut {NoStop}%
		\bibitem [{\citenamefont {Singh}\ \emph {et~al.}(2018)\citenamefont {Singh},
			\citenamefont {Soulages},\ and\ \citenamefont {Ewoldt}}]{Singh2018}%
		\BibitemOpen
		\bibfield  {author} {\bibinfo {author} {\bibfnamefont {P.~K.}\ \bibnamefont
				{Singh}}, \bibinfo {author} {\bibfnamefont {J.~M.}\ \bibnamefont {Soulages}},
			\ and\ \bibinfo {author} {\bibfnamefont {R.~H.}\ \bibnamefont {Ewoldt}},\
		}\href {\doibase 10.1122/1.4999795} {\bibfield  {journal} {\bibinfo
				{journal} {Journal of Rheology}\ }\textbf {\bibinfo {volume} {62}},\ \bibinfo
			{pages} {277} (\bibinfo {year} {2018})}\BibitemShut {NoStop}%
		\bibitem [{\citenamefont {Hyun}\ and\ \citenamefont
			{Wilhelm}(2008)}]{Hyun2008}%
		\BibitemOpen
		\bibfield  {author} {\bibinfo {author} {\bibfnamefont {K.}~\bibnamefont
				{Hyun}}\ and\ \bibinfo {author} {\bibfnamefont {M.}~\bibnamefont {Wilhelm}},\
		}\href {\doibase 10.1063/1.2964694} {\bibfield  {journal} {\bibinfo
				{journal} {AIP Conference Proceedings}\ }\textbf {\bibinfo {volume} {1027}},\
			\bibinfo {pages} {369} (\bibinfo {year} {2008})}\BibitemShut {NoStop}%
		\bibitem [{\citenamefont {Song}\ and\ \citenamefont {Hyun}(2018)}]{Song2018}%
		\BibitemOpen
		\bibfield  {author} {\bibinfo {author} {\bibfnamefont {H.~Y.}\ \bibnamefont
				{Song}}\ and\ \bibinfo {author} {\bibfnamefont {K.}~\bibnamefont {Hyun}},\
		}\href {\doibase 10.1122/1.5024720} {\bibfield  {journal} {\bibinfo
				{journal} {Journal of Rheology}\ }\textbf {\bibinfo {volume} {62}},\ \bibinfo
			{pages} {919} (\bibinfo {year} {2018})}\BibitemShut {NoStop}%
		\bibitem [{\citenamefont {Song}\ and\ \citenamefont {Hyun}(2019)}]{Song2019}%
		\BibitemOpen
		\bibfield  {author} {\bibinfo {author} {\bibfnamefont {H.~Y.}\ \bibnamefont
				{Song}}\ and\ \bibinfo {author} {\bibfnamefont {K.}~\bibnamefont {Hyun}},\
		}\href {\doibase 10.1007/s13367-019-0001-x} {\bibfield  {journal} {\bibinfo
				{journal} {Korea-Australia Rheology Journal}\ }\textbf {\bibinfo {volume}
				{31}},\ \bibinfo {pages} {1} (\bibinfo {year} {2019})}\BibitemShut {NoStop}%
		\bibitem [{\citenamefont {Carey-De La~Torre}\ and\ \citenamefont
			{Ewoldt}(2018)}]{Carey-DeLaTorre2018}%
		\BibitemOpen
		\bibfield  {author} {\bibinfo {author} {\bibfnamefont {O.}~\bibnamefont
				{Carey-De La~Torre}}\ and\ \bibinfo {author} {\bibfnamefont {R.~H.}\
				\bibnamefont {Ewoldt}},\ }\href {\doibase 10.1007/s13367-018-0001-2}
		{\bibfield  {journal} {\bibinfo  {journal} {Korea-Australia Rheology
					Journal}\ }\textbf {\bibinfo {volume} {30}},\ \bibinfo {pages} {1} (\bibinfo
			{year} {2018})}\BibitemShut {NoStop}%
		\bibitem [{\citenamefont {Hyun}\ \emph {et~al.}(2007)\citenamefont {Hyun},
			\citenamefont {Baik}, \citenamefont {Ahn}, \citenamefont {Lee}, \citenamefont
			{Sugimoto},\ and\ \citenamefont {Koyama}}]{Hyun2007}%
		\BibitemOpen
		\bibfield  {author} {\bibinfo {author} {\bibfnamefont {K.}~\bibnamefont
				{Hyun}}, \bibinfo {author} {\bibfnamefont {E.~S.}\ \bibnamefont {Baik}},
			\bibinfo {author} {\bibfnamefont {K.~H.}\ \bibnamefont {Ahn}}, \bibinfo
			{author} {\bibfnamefont {S.~J.}\ \bibnamefont {Lee}}, \bibinfo {author}
			{\bibfnamefont {M.}~\bibnamefont {Sugimoto}}, \ and\ \bibinfo {author}
			{\bibfnamefont {K.}~\bibnamefont {Koyama}},\ }\href {\doibase
			10.1122/1.2790072} {\bibfield  {journal} {\bibinfo  {journal} {Journal of
					Rheology}\ }\textbf {\bibinfo {volume} {51}},\ \bibinfo {pages} {1319}
			(\bibinfo {year} {2007})}\BibitemShut {NoStop}%
\bibitem [{\citenamefont {Mykhaylyk}\ \emph {et~al.}(2011)\citenamefont
	{Mykhaylyk}, \citenamefont {Fernyhough}, \citenamefont {Okura}, \citenamefont
	{Fairclough}, \citenamefont {Ryan},\ and\ \citenamefont
	{Graham}}]{Mykhalyk2011}%
\BibitemOpen
\bibfield  {author} {\bibinfo {author} {\bibfnamefont {O.~O.}\ \bibnamefont
		{Mykhaylyk}}, \bibinfo {author} {\bibfnamefont {C.~M.}\ \bibnamefont
		{Fernyhough}}, \bibinfo {author} {\bibfnamefont {M.}~\bibnamefont {Okura}},
	\bibinfo {author} {\bibfnamefont {J.~P.~A.}\ \bibnamefont {Fairclough}},
	\bibinfo {author} {\bibfnamefont {A.~J.}\ \bibnamefont {Ryan}}, \ and\
	\bibinfo {author} {\bibfnamefont {R.}~\bibnamefont {Graham}},\ }\href
{\doibase https://doi.org/10.1016/j.eurpolymj.2010.09.021} {\bibfield
	{journal} {\bibinfo  {journal} {European Polymer Journal}\ }\textbf {\bibinfo
		{volume} {47}},\ \bibinfo {pages} {447 } (\bibinfo {year}
	{2011})}\BibitemShut {NoStop}%
		\bibitem [{\citenamefont {Ferry}(1980)}]{Ferry1980}%
		\BibitemOpen
		\bibfield  {author} {\bibinfo {author} {\bibfnamefont {J.~D.}\ \bibnamefont
				{Ferry}},\ }\href@noop {} {\emph {\bibinfo {title} {{Viscoelastic Properties
						of Polymers, 3rd Edition}}}}\ (\bibinfo  {publisher} {Wiley, John {\&} Sons,
			Incorporated (1987)},\ \bibinfo {address} {New York},\ \bibinfo {year}
		{1980})\ p.\ \bibinfo {pages} {672}\BibitemShut {NoStop}%
		\bibitem [{Note1()}]{Note1}%
		\BibitemOpen
		\bibinfo {note} {\protect \url {http://www.reptate.com}}\BibitemShut
		{NoStop}%
		\bibitem [{Note2()}]{Note2}%
		\BibitemOpen
		\bibinfo {note} {\protect \url
			{https://sourceforge.net/projects/cdrheo/}}\BibitemShut {NoStop}%
		\bibitem [{\citenamefont {S.~Poulos}\ \emph {et~al.}(2015)\citenamefont
			{S.~Poulos}, \citenamefont {Renou}, \citenamefont {R.~Jacob}, \citenamefont
			{Koumakis},\ and\ \citenamefont {Petekidis}}]{Poulos2015}%
		\BibitemOpen
		\bibfield  {author} {\bibinfo {author} {\bibfnamefont {A.}~\bibnamefont
				{S.~Poulos}}, \bibinfo {author} {\bibfnamefont {F.}~\bibnamefont {Renou}},
			\bibinfo {author} {\bibfnamefont {A.}~\bibnamefont {R.~Jacob}}, \bibinfo
			{author} {\bibfnamefont {N.}~\bibnamefont {Koumakis}}, \ and\ \bibinfo
			{author} {\bibfnamefont {G.}~\bibnamefont {Petekidis}},\ }\href {\doibase
			10.1007/s00397-015-0865-8} {\bibfield  {journal} {\bibinfo  {journal}
				{Rheologica Acta}\ }\textbf {\bibinfo {volume} {accepted}} (\bibinfo {year}
			{2015}),\ 10.1007/s00397-015-0865-8}\BibitemShut {NoStop}%
		\bibitem [{\citenamefont {Reynolds}(2018)}]{Reynolds2018a}%
		\BibitemOpen
		\bibfield  {author} {\bibinfo {author} {\bibfnamefont {C.~D.}\ \bibnamefont
				{Reynolds}},\ }\emph {\bibinfo {title} {{Rheological behaviour of polymer
					melts and its relationship with underlying structure and topology}}},\ \href
		{http://etheses.dur.ac.uk/12665/} {Ph.D. thesis},\ \bibinfo  {school} {Durham
			University} (\bibinfo {year} {2018})\BibitemShut {NoStop}%
		\bibitem [{\citenamefont {Likhtman}\ and\ \citenamefont
			{McLeish}(2002)}]{Likhtman2002}%
		\BibitemOpen
		\bibfield  {author} {\bibinfo {author} {\bibfnamefont {A.~E.}\ \bibnamefont
				{Likhtman}}\ and\ \bibinfo {author} {\bibfnamefont {T.~C.~B.}\ \bibnamefont
				{McLeish}},\ }\href {\doibase 10.1021/ma0200219} {\bibfield  {journal}
			{\bibinfo  {journal} {Macromolecules}\ }\textbf {\bibinfo {volume} {35}},\
			\bibinfo {pages} {6332} (\bibinfo {year} {2002})}\BibitemShut {NoStop}%
		\bibitem [{\citenamefont {Graham}\ \emph {et~al.}(2003)\citenamefont {Graham},
			\citenamefont {Likhtman}, \citenamefont {McLeish},\ and\ \citenamefont
			{Milner}}]{Graham2003}%
		\BibitemOpen
		\bibfield  {author} {\bibinfo {author} {\bibfnamefont {R.~S.}\ \bibnamefont
				{Graham}}, \bibinfo {author} {\bibfnamefont {A.~E.}\ \bibnamefont
				{Likhtman}}, \bibinfo {author} {\bibfnamefont {T.~C.~B.}\ \bibnamefont
				{McLeish}}, \ and\ \bibinfo {author} {\bibfnamefont {S.~T.}\ \bibnamefont
				{Milner}},\ }\href {\doibase 10.1122/1.1595099} {\bibfield  {journal}
			{\bibinfo  {journal} {Journal of Rheology}\ }\textbf {\bibinfo {volume}
				{47}},\ \bibinfo {pages} {1171} (\bibinfo {year} {2003})}\BibitemShut
		{NoStop}%
		\bibitem [{\citenamefont {Auhl}\ \emph {et~al.}(2008)\citenamefont {Auhl},
			\citenamefont {Ramirez}, \citenamefont {Likhtman}, \citenamefont {Chambon},\
			and\ \citenamefont {Fernyhough}}]{Auhl2008a}%
		\BibitemOpen
		\bibfield  {author} {\bibinfo {author} {\bibfnamefont {D.~W.}\ \bibnamefont
				{Auhl}}, \bibinfo {author} {\bibfnamefont {J.}~\bibnamefont {Ramirez}},
			\bibinfo {author} {\bibfnamefont {A.~E.}\ \bibnamefont {Likhtman}}, \bibinfo
			{author} {\bibfnamefont {P.}~\bibnamefont {Chambon}}, \ and\ \bibinfo
			{author} {\bibfnamefont {C.}~\bibnamefont {Fernyhough}},\ }\href {\doibase
			10.1122/1.2890780} {\bibfield  {journal} {\bibinfo  {journal} {Journal of
					Rheology}\ }\textbf {\bibinfo {volume} {52}},\ \bibinfo {pages} {801}
			(\bibinfo {year} {2008})}\BibitemShut {NoStop}%
		\bibitem [{\citenamefont {Likhtman}\ and\ \citenamefont
			{Graham}(2003)}]{Likhtman2003}%
		\BibitemOpen
		\bibfield  {author} {\bibinfo {author} {\bibfnamefont {A.~E.}\ \bibnamefont
				{Likhtman}}\ and\ \bibinfo {author} {\bibfnamefont {R.~S.}\ \bibnamefont
				{Graham}},\ }\href {\doibase 10.1016/S0377-0257(03)00114-9} {\bibfield
			{journal} {\bibinfo  {journal} {Journal of Non-Newtonian Fluid Mechanics}\
			}\textbf {\bibinfo {volume} {114}},\ \bibinfo {pages} {1} (\bibinfo {year}
			{2003})}\BibitemShut {NoStop}%
		\bibitem [{\citenamefont {Hoyle}(2010)}]{Hoyle2010a}%
		\BibitemOpen
		\bibfield  {author} {\bibinfo {author} {\bibfnamefont {D.~M.}\ \bibnamefont
				{Hoyle}},\ }\emph {\bibinfo {title} {{Constitutive modelling of branched
					polymer melts in non-linear response}}},\ \href
		{http://etheses.whiterose.ac.uk/id/eprint/1434} {Ph.D. thesis},\ \bibinfo
		{school} {University of Leeds} (\bibinfo {year} {2010})\BibitemShut {NoStop}%
	\end{thebibliography}

\clearpage

\begin{thebibliography}{44}%
	\makeatletter
	\providecommand \@ifxundefined [1]{%
		\@ifx{#1\undefined}
	}%
	\providecommand \@ifnum [1]{%
		\ifnum #1\expandafter \@firstoftwo
		\else \expandafter \@secondoftwo
		\fi
	}%
	\providecommand \@ifx [1]{%
		\ifx #1\expandafter \@firstoftwo
		\else \expandafter \@secondoftwo
		\fi
	}%
	\providecommand \natexlab [1]{#1}%
	\providecommand \enquote  [1]{``#1''}%
	\providecommand \bibnamefont  [1]{#1}%
	\providecommand \bibfnamefont [1]{#1}%
	\providecommand \citenamefont [1]{#1}%
	\providecommand \href@noop [0]{\@secondoftwo}%
	\providecommand \href [0]{\begingroup \@sanitize@url \@href}%
	\providecommand \@href[1]{\@@startlink{#1}\@@href}%
	\providecommand \@@href[1]{\endgroup#1\@@endlink}%
	\providecommand \@sanitize@url [0]{\catcode `\\12\catcode `\$12\catcode
		`\&12\catcode `\#12\catcode `\^12\catcode `\_12\catcode `\%12\relax}%
	\providecommand \@@startlink[1]{}%
	\providecommand \@@endlink[0]{}%
	\providecommand \url  [0]{\begingroup\@sanitize@url \@url }%
	\providecommand \@url [1]{\endgroup\@href {#1}{\urlprefix }}%
	\providecommand \urlprefix  [0]{URL }%
	\providecommand \Eprint [0]{\href }%
	\providecommand \doibase [0]{http://dx.doi.org/}%
	\providecommand \selectlanguage [0]{\@gobble}%
	\providecommand \bibinfo  [0]{\@secondoftwo}%
	\providecommand \bibfield  [0]{\@secondoftwo}%
	\providecommand \translation [1]{[#1]}%
	\providecommand \BibitemOpen [0]{}%
	\providecommand \bibitemStop [0]{}%
	\providecommand \bibitemNoStop [0]{.\EOS\space}%
	\providecommand \EOS [0]{\spacefactor3000\relax}%
	\providecommand \BibitemShut  [1]{\csname bibitem#1\endcsname}%
	\let\auto@bib@innerbib\@empty
	
	\bibitem [{\citenamefont {Wilhelm}(2002)}]{Wilhelm2002-SI}%
	\BibitemOpen
	\bibfield  {author} {\bibinfo {author} {\bibfnamefont {M.}~\bibnamefont
			{Wilhelm}},\ }\href {\doibase
		10.1002/1439-2054(20020201)287:2<83::AID-MAME83>3.0.CO;2-B} {\bibfield
		{journal} {\bibinfo  {journal} {Macromolecular Materials and Engineering}\
		}\textbf {\bibinfo {volume} {287}},\ \bibinfo {pages} {83} (\bibinfo {year}
		{2002})}\BibitemShut {NoStop}%
	\bibitem [{\citenamefont {S.~Poulos}\ \emph {et~al.}(2015)\citenamefont
		{S.~Poulos}, \citenamefont {Renou}, \citenamefont {R.~Jacob}, \citenamefont
		{Koumakis},\ and\ \citenamefont {Petekidis}}]{Poulos2015-SI}%
	\BibitemOpen
	\bibfield  {author} {\bibinfo {author} {\bibfnamefont {A.}~\bibnamefont
			{S.~Poulos}}, \bibinfo {author} {\bibfnamefont {F.}~\bibnamefont {Renou}},
		\bibinfo {author} {\bibfnamefont {A.}~\bibnamefont {R.~Jacob}}, \bibinfo
		{author} {\bibfnamefont {N.}~\bibnamefont {Koumakis}}, \ and\ \bibinfo
		{author} {\bibfnamefont {G.}~\bibnamefont {Petekidis}},\ }\href {\doibase
		10.1007/s00397-015-0865-8} {\bibfield  {journal} {\bibinfo  {journal}
			{Rheologica Acta}\ }\textbf {\bibinfo {volume} {accepted}} (\bibinfo {year}
		{2015}),\ 10.1007/s00397-015-0865-8}\BibitemShut {NoStop}%
	\bibitem [{\citenamefont {Hoyle}\ \emph {et~al.}(2014)\citenamefont {Hoyle},
		\citenamefont {Auhl}, \citenamefont {Harlen}, \citenamefont {Barroso},
		\citenamefont {Wilhelm},\ and\ \citenamefont {McLeish}}]{Hoyle2014-SI}%
	\BibitemOpen
	\bibfield  {author} {\bibinfo {author} {\bibfnamefont {D.~M.}\ \bibnamefont
			{Hoyle}}, \bibinfo {author} {\bibfnamefont {D.~W.}\ \bibnamefont {Auhl}},
		\bibinfo {author} {\bibfnamefont {O.~G.}\ \bibnamefont {Harlen}}, \bibinfo
		{author} {\bibfnamefont {V.~C.}\ \bibnamefont {Barroso}}, \bibinfo {author}
		{\bibfnamefont {M.}~\bibnamefont {Wilhelm}}, \ and\ \bibinfo {author}
		{\bibfnamefont {T.~C.~B.}\ \bibnamefont {McLeish}},\ }\href {\doibase
		10.1122/1.4881467} {\bibfield  {journal} {\bibinfo  {journal} {Journal of
				Rheology}\ }\textbf {\bibinfo {volume} {58}},\ \bibinfo {pages} {969}
		(\bibinfo {year} {2014})}\BibitemShut {NoStop}%
	\bibitem [{\citenamefont {Likhtman}\ and\ \citenamefont
		{McLeish}(2002)}]{Likhtman2002-SI}%
	\BibitemOpen
	\bibfield  {author} {\bibinfo {author} {\bibfnamefont {A.~E.}\ \bibnamefont
			{Likhtman}}\ and\ \bibinfo {author} {\bibfnamefont {T.~C.~B.}\ \bibnamefont
			{McLeish}},\ }\href {\doibase 10.1021/ma0200219} {\bibfield  {journal}
		{\bibinfo  {journal} {Macromolecules}\ }\textbf {\bibinfo {volume} {35}},\
		\bibinfo {pages} {6332} (\bibinfo {year} {2002})}\BibitemShut {NoStop}%
	\bibitem [{\citenamefont {Graham}\ \emph {et~al.}(2003)\citenamefont {Graham},
		\citenamefont {Likhtman}, \citenamefont {McLeish},\ and\ \citenamefont
		{Milner}}]{Graham2003-SI}%
	\BibitemOpen
	\bibfield  {author} {\bibinfo {author} {\bibfnamefont {R.~S.}\ \bibnamefont
			{Graham}}, \bibinfo {author} {\bibfnamefont {A.~E.}\ \bibnamefont
			{Likhtman}}, \bibinfo {author} {\bibfnamefont {T.~C.~B.}\ \bibnamefont
			{McLeish}}, \ and\ \bibinfo {author} {\bibfnamefont {S.~T.}\ \bibnamefont
			{Milner}},\ }\href {\doibase 10.1122/1.1595099} {\bibfield  {journal}
		{\bibinfo  {journal} {Journal of Rheology}\ }\textbf {\bibinfo {volume}
			{47}},\ \bibinfo {pages} {1171} (\bibinfo {year} {2003})}\BibitemShut
	{NoStop}%
	\bibitem [{\citenamefont {Auhl}\ \emph {et~al.}(2008)\citenamefont {Auhl},
		\citenamefont {Ramirez}, \citenamefont {Likhtman}, \citenamefont {Chambon},\
		and\ \citenamefont {Fernyhough}}]{Auhl2008a-SI}%
	\BibitemOpen
	\bibfield  {author} {\bibinfo {author} {\bibfnamefont {D.~W.}\ \bibnamefont
			{Auhl}}, \bibinfo {author} {\bibfnamefont {J.}~\bibnamefont {Ramirez}},
		\bibinfo {author} {\bibfnamefont {A.~E.}\ \bibnamefont {Likhtman}}, \bibinfo
		{author} {\bibfnamefont {P.}~\bibnamefont {Chambon}}, \ and\ \bibinfo
		{author} {\bibfnamefont {C.}~\bibnamefont {Fernyhough}},\ }\href {\doibase
		10.1122/1.2890780} {\bibfield  {journal} {\bibinfo  {journal} {Journal of
				Rheology}\ }\textbf {\bibinfo {volume} {52}},\ \bibinfo {pages} {801}
		(\bibinfo {year} {2008})}\BibitemShut {NoStop}%
	\bibitem [{\citenamefont {Hoyle}(2010)}]{Hoyle2010a-SI}%
	\BibitemOpen
	\bibfield  {author} {\bibinfo {author} {\bibfnamefont {D.~M.}\ \bibnamefont
			{Hoyle}},\ }\emph {\bibinfo {title} {{Constitutive modelling of branched
				polymer melts in non-linear response}}},\ \href
	{http://etheses.whiterose.ac.uk/id/eprint/1434} {Ph.D. thesis},\ \bibinfo
	{school} {University of Leeds} (\bibinfo {year} {2010})\BibitemShut {NoStop}%
	\bibitem [{\citenamefont {Gurnon}(2012)}]{Gurnon2012-SI}%
	\BibitemOpen
	\bibfield  {author} {\bibinfo {author} {\bibfnamefont {M.}~\bibnamefont
			{Gurnon}},\ }\href {\doibase
		10.1122/1.3684751} {\bibfield
		{journal} {\bibinfo  {journal} {Journal of Rheology}\
		}\textbf {\bibinfo {volume} {56}}, \bibinfo {number} {2}, \bibinfo {pages} {333-351} (\bibinfo {year}
		{2012})}\BibitemShut {NoStop}%
	\bibitem [{\citenamefont {Calin}(2010)}]{Calin2010-SI}%
	\BibitemOpen
	\bibfield  {author} {\bibinfo {author} {\bibfnamefont {M.}~\bibnamefont
			{Calin}},\ }\href {\doibase
		10.1122/1.3684751} {\bibfield
		{journal} {\bibinfo  {journal} {Journal of Non-Newtonian Fluid Mechanics}\
		}\textbf {\bibinfo {volume} {165}}, \bibinfo {number} {23}, \bibinfo {pages} {1564 - 1577} (\bibinfo {year}
		{2010})}\BibitemShut {NoStop}%
\end{thebibliography}
\end{document}